\RequirePackage{ifpdf}
\documentclass[12pt,a4paper,hyper]{JHEP3}

\usepackage{epsfig}
\usepackage{amsmath}
\usepackage{multirow}
\usepackage{epsfig}
\usepackage{psfrag}
\usepackage{color}

\def\={\;  = \;}

\title{\center {Nonperturbative Black Hole Entropy  \\ and  Kloosterman Sums}}

\preprint{}
\author{
Atish Dabholkar$^{1, 2}$, Jo\~ao Gomes$^{3}$, and Sameer Murthy$^{4}$\\

\it $^1${Sorbonne Universit\'es, UPMC Univ Paris 06\\
\it  UMR 7589, LPTHE, F-75005, Paris, France}\\

\it $^2${CNRS, UMR 7589, LPTHE, F-75005, Paris, France}\\

\it $^3$Department of Applied Mathematics and Theoretical Physics\\
\it University of Cambridge, 
\it Wilberforce road, CB3 0WA,  UK \\

\it $^4$Department of Mathematics, King's College\\
The Strand, WC2R 2LS, London, UK\\

{\rm Emails : atish at lpthe.jussieu.fr, {jmg84 at cam.ac.uk},
{sameer dot murthy at kcl.ac.uk}}\\
}

\abstract{Non-perturbative quantum corrections to supersymmetric black hole entropy often involve nontrivial  number-theoretic phases called Kloosterman sums. We show how these sums can be obtained  naturally from the functional integral of supergravity in asymptotically $AdS_{2}$ space for a  class of black holes. They are essentially topological in origin and correspond to  charge-dependent phases arising  from the  various   gauge and gravitational Chern-Simons terms and boundary Wilson lines evaluated on Dehn-filled solid 2-torus. These corrections are essential to obtain an integer from supergravity in agreement with the quantum degeneracies, and reveal an intriguing connection between topology, number theory, and quantum gravity.  We give an assessment of the current understanding of quantum  entropy of  black holes.}

\keywords{black holes, superstrings, holography}

\newenvironment{myenumerate}{
\begin{enumerate}
   \setlength{\itemsep}{1pt}
   \setlength{\parskip}{0pt}
   \setlength{\parsep}{0pt}}{\end{enumerate}}

\newcommand{\IZ}{\mathbb{Z}}

\newcommand{\bem}{\begin{pmatrix}}
\newcommand{\eem}{\end{pmatrix}}

\def\vth{\vartheta}

\def\r{\rho}

\def\g{\gamma}
\def\t{\tau}
\def\a{\alpha}
\def\b{\beta}
\def\m{\mu}
\def\n{\nu}
\def\e{\epsilon}

\def\h{\eta}
\def\l{{\lambda}}

\def\O{{\Omega}}
\def\D{\Delta}

\def\CN{{\cal N}}

\def\half{{\frac12}}

\def\CN{{\cal N}}

\def\bea{\begin{eqnarray}}
\def\eea{\end{eqnarray}}
\def\be{\begin{equation}}
\def\ee{\end{equation}}
\def\ba{\begin{align}}
\def\ea{\end{align}}
\def\bse{\begin{subequations}}
\def\ese{\end{subequations}}
\def\1F1{{}_1\!F_1}
\def\2F0{{}_2\!F_0}

\def\a{\alpha}

\def\h3{$\textrm{H}_3^+$}

\def\IZ{{\mathbb Z}}

\newcommand{\beq}{\begin{equation}}
\newcommand{\eeq}{\end{equation}}
\newcommand{\ber}{\begin{eqnarray}}
\newcommand{\eer}{\end{eqnarray}}

\def\be{\begin{eqnarray}}
\def\ee{\end{eqnarray}}

\newcommand{\cO}{{\cal O}}

\def\wh{\widehat}

\def\mod{{\rm mod}}

\def\CN{{\cal N}}

\font\manual=manfnt
\def\dbend{\lower3.5pt\hbox{\manual\char127}}

\def\c{\cdot}

\def\bar{\overline}

\def\CN{{\cal N}}

\def\rt2{\sqrt{2}}
\def\irt2{{1\over\sqrt{2}}}

\def\wt{\widetilde}

\def\b{\beta}
\def\a{\alpha}

\def\g{\gamma}

\def\mod{{\rm mod}}

\font\cmss=cmss10
\font\cmsss=cmss10 at 7pt

\def\IL{\relax{\rm I\kern-.18em L}}
\def\IH{\relax{\rm I\kern-.18em H}}
\def\rlx{\relax\leavevmode}
\def\ZZ{\rlx\leavevmode\ifmmode\mathchoice{\hbox{\cmss Z\kern-.4em Z}}
 {\hbox{\cmss Z\kern-.4em Z}}{\lower.9pt\hbox{\cmsss Z\kern-.36em Z}}
 {\lower1.2pt\hbox{\cmsss Z\kern-.36em Z}}\else{\cmss Z\kern-.4em
 Z}\fi}

\begin{document}

\section{Introduction}

Quantum entropy of a black hole is the full quantum generalization of the celebrated Bekenstein-Hawking entropy including  both perturbative and nonperturbative corrections. For supersymmetric black holes, it is defined by a formal functional integral of massless supergravity fields in the near horizon $AdS_{2}$ space in the framework of $AdS_{2}/CFT_{1}$ correspondence~\cite{Sen:2008vm,Sen:2008yk}. The measure of the functional integral is determined by  the effective Wilsonian action including all higher derivative terms obtained by integrating out the massive string fields.  

There has been considerable progress in evaluating this  functional integral for  a class of supersymmetric black holes using localization techniques~\cite{Dabholkar:2010uh, Dabholkar:2011ec, Banerjee:2009af}. In particular, in the simple example of one-eighth BPS black holes in $\CN=8$ supersymmetric theory in four dimensions, it seems possible to exactly sum up all perturbative  corrections to the Bekenstein-Hawking area formula~\cite{Dabholkar:2011ec}. It is also clear qualitatively that the nonperturbative corrections expected from the microscopic degeneracy can be reproduced from  contributions from subleading saddle points obtained by smooth  $\IZ_{c}$ orbifolds of  $AdS_{2}$ labeled by a positive integer $c$~\cite{Banerjee:2008ky, Murthy:2009dq}.

Our goal here is to  compute explicitly the  nonperturbative corrections which involve subtle number theoretic phases called the Kloosterman sums. It has been a long standing puzzle how these phases could possibly arise from a supergravity functional integral.  These corrections,  even though subleading, are conceptually very important  because the exponential of the quantum entropy is expected to yield precisely an integer equal to the quantum degeneracy of the black hole. The nonpertubative corrections including the precise Kloosterman sums are essential to obtain this integrality. 
This is a nontrivial and stringent constraint. Indeed, it is something of a surprise that a functional integral in the bulk which is intrinsically a complex analytic continuous object could yield  a specific integer which is a number theoretic discrete object. 

We emphasize that it is meaningful to discuss  nonperturbative corrections arising from subleading saddle points only because the full perturbative answer around the leading saddle point can be evaluated \textit{exactly} using localization. 
In a typical physics context, when a  functional integral is evaluated in the saddle point approximation,  the perturbative expansion   around the saddle point involves all loops.  Usually there is no practical way to evaluate the higher loops. Even if one could, it is  only an asymptotic expansion and not a convergent one. So, it is not particularly meaningful to include the contribution from subleading saddle points which are exponentially smaller compared to the  perturbative contributions around the leading saddle point. In the present context, by contrast, the  evaluation of the  functional integral around the localizing saddle point is one-loop exact and can be evaluated explicitly. 
 
Somewhat surprisingly, the supergravity functional integral seems capable of reproducing the intricate structure of the Kloosterman sums. It turns out to be essentially topological in origin. The supergravity action includes Chern-Simons terms (including the higher derivative gravitational Chern-Simons terms) and  the quantum entropy functional integral includes boundary Wilson lines. These terms evaluated on the subleading localizing instantons  lead to charge-dependent phases which combine nontrivially to yield precisely the Kloosterman sums.  

Quantum Entropy connects to the broader problem of Quantum Holography, that is,  of understanding finite $N$ quantum effects in the bulk quantum gravity in $AdS/CFT$ holography. Developing on our results can provide the first example of quantum holography where both the bulk and boundary partition functions are computable including all quantum corrections.  

The plan of the paper is as follows.  In \S\ref{Klooster} we review the definition of the classical and generalized Kloosterman sums and the associated Rademacher expansions.
In \S\ref{Local} we discuss the $SL(2, \mathbb{Z})$ family of  localizing orbifold solutions that contribute to the nonperturbative corrections to the quantum entropy.  Sections  \S\ref{Phases} and \S\ref{Multiplier} contain the main results of this paper where we describe how various phases and the multiplier system  in the Kloosterman sums are reproduced from  the orbifolded localization instantons. In \S\ref{Assess} we give an assessment of  the  hurdles that have been overcome in this program and  the problems that remain to be solved. 

\section{Kloosterman Sums and the Rademacher Expansion \label{Klooster}}

We review the definitions  of Kloosterman sums and their relation to the Rademacher Expansion.

\subsection{The Classical Kloosterman Sum}

The classical Kloosterman sum for integers $n, m, c$ is defined by
\begin{equation}\label{classical Kl}
 Kl(n, m, c) :=\sum_{\substack{ d\in \mathbb{Z}/c\mathbb{Z}\\
                  d a =1\,\text{mod}(c)}}e^{2\pi i \left(n \frac{d}{c}+ m \frac{a}{c}\right)}.
\end{equation}

It arises naturally in analytic number theory in the study of Fourier coefficients of modular forms of negative weight.
A modular form of particular interest to us (see \S\ref{Micro}) is 
\be \label{half-BPS}
F(\tau) = \frac{1}{q}  \prod_{n=1}^{\infty} \frac{1}{ (1-q^n)^{24}} \,  \qquad  (q:= e^{2\pi i \tau})\, ,
\ee  
which  has weight $-12$:
\be
F\left(\frac{a \tau +b}{c\tau +d}\right) = (c\tau +d )^{-12} F (\tau) \,  
\ee
with Fourier expansion
\begin{equation}
F (\tau )=\sum_{N=-1}^{\infty} c(N)q^{N}=q^{-1}+24+\ldots \, \, .
\end{equation}
The modular properties of  $F(\tau)$ and the fact that it has negative weight imply that $c(N)$ admits the  Hardy-Ramanujan-Rademacher expansion for $N \geq 0$:
\begin{equation}\label{kloos1}
c(N)=\sum_{c=1}^{\infty} \left(\frac{2\pi}{c}\right)^{14}Kl(N, -1, c) \,  {I}_{13}\left(\frac{4\pi\sqrt{N}}{c}\right) \, , 
\end{equation} 
where 
\begin{equation}\label{intrep}
 I_{\rho}(z) :=\frac{1}{2\pi i}\int_{\epsilon-i\infty}^{\epsilon+i\infty} \, \frac{dt}{t^{\r +1}}\exp [{t +\frac{z^2}{4t}}]
\, 
\end{equation}
is the modified Bessel function of index $\r$.  The Kloosterman sum \eqref{classical Kl} can be regarded as a discrete analog of this integral representation of the Bessel function\footnote{We thank E. Witten for this observation.}  where  $d$  is the analog of the integration variable $t$, and  $a = d^{*}$ is the analog\footnote{The inverse  of  an integer $x$ in the finite field $\mathbb{Z}/c\mathbb{Z}$  is usually denoted by   $x^*$, thus $a = d^{*}$ as $ad =1\, \mod\, c$.}  of $t^{-1}$.  

For any integer $p$ coprime to $c$, the Kloosterman sum satisfies 
\be\label{kloosrel}
Kl(np, m,c)=Kl(n, mp, c)
\ee
which can be proven easily by changing variables $pd=\tilde{d}$ and $d^*=p\tilde{d}^*$. This property will be important later in \S\ref{Sec grav CS} to prove the   duality invariance of the bulk answers.

\subsection{The Generalized Kloosterman Sum}

There is a natural generalization of  the Rademacher expansion for  Fourier coefficients  of weak Jacobi forms, which suggests the following definition for the generalized Kloosterman sum.  For a review of  weak Jacobi forms see, for example,~\cite{Eichler:1985ja} or~\cite{Dabholkar:2012nd}. 
    
Consider the standard weight-half index-$k$ theta functions defined by\footnote{In \cite{Eichler:1985ja, Dabholkar:2012nd},  the weight  is denoted by $k$ and  the index is denoted by $m$. We  follow the physics conventions here to denote  the index  by $k$ because it will later be related to a Chern-Simons level usually denoted by $k$.} 
\begin{equation}
 \vartheta_{k,\mu}(\tau,z)=\sum_{\substack{l\in \mathbb{Z} \\ l=\mu \, \text{mod} \, 2k}} q^{l^2/{4k}}y^l \, , \qquad q:= e^{2\pi i \tau} \, , \, y:= e^{2\pi i z} 
\end{equation} 
for a positive integer $k$. 
Under  modular transformation they transform among each other  as \begin{equation}\label{thetadef}
 \vartheta_{k,\nu}\left(\frac{a \tau +b}{c\tau +d}, \frac{z}{c\tau+d}\right) \, = \, \sqrt{(c\tau+d)} \, e^{2\pi i k \frac{cz^2}{c\tau+d}} \, M^{-1}(\gamma)_{\nu \mu} \, \vartheta_{k,\mu}(\tau,z) \, , 
\end{equation} 
where
\be
 \gamma = \left(\begin{array}{cc}a & b \\c & d\end{array}\right)
\in SL(2,\mathbb{Z}) \, ,
\ee 
and the square root is defined by choosing the principal branch of the logarithm $\log{z}=\log|z|+\arg{z},\,-\pi<\arg{z}\leq\pi$.
The matrix $M(\gamma)$ is called the  multiplier system. To obtain an explicit analytic expression for the representation $M(\g)$ is a subtle problem in number theory. 
In  \S \ref{multiplier sect}, we review a  representation obtained in \cite{Lisa92} that   is particularly well-suited  for  our physical computations in \S\ref{Physical} because it affords a natural path integral interpretation \cite{Witten:1988hf}.

For a given $k$, the generalized Kloosterman sum \cite{Rademacher:1964ra,Dijkgraaf:2000fq,Manschot:2007ha} is then defined by
\begin{equation}\label{gener Kloosterman}
 Kl(n, m, c; \nu, \mu):= \sum_{\substack{-c< d \leq 0 \\ (c,d)=1}}e^{2\pi i(\frac{\Delta}{4 k})\frac{d}{c}} \, M^{-1}(\gamma)_{\nu \mu} \, e^{2\pi i(\frac{\tilde \Delta}{4 k})\frac{a}{c}}\, ,
 \end{equation}
where $\Delta := 4 k n - \nu^{2}$ and $\tilde \Delta  := 4 k m  - \mu^{2}$.  The summand depends only on the equivalence class 
 $[\gamma]\in \Gamma_{\infty}\backslash SL(2,\mathbb{Z})/\Gamma_{\infty}$ where $\Gamma_{\infty}$ is the subgroup of translations generated by $T$. This can be checked  using the definition   \eqref{thetadef} of $M ^{{-1}}(\gamma)_{\mu\nu}$. Practically, this means that  by the left and the right coset action we can  choose a representative of the  class $[\gamma]$ such that $-c <a, d \leq 0$. 

The generalized Kloosterman sum appears naturally in the Rademacher expansion of Fourier coefficients  of  weak Jacobi forms.
Consider a weak Jacobi form $ \phi(\tau,z)$ of nonpositive weight  $w$ and index $k$.  Elliptic properties of the Jacobi form imply a  theta expansion:
\begin{equation}
 \phi(\tau,z)=\sum_{-k+1\leq \nu\leq k} h_{\nu}(\tau) \, \vartheta_{k,\nu}(\tau,z)
\end{equation}
where  $h_{\nu}(\tau)$ is  periodic in $\nu$ with period $2k$. Modular properties of the Jacobi form imply that the components of the vector   transform under a modular transformation as
\begin{equation}
 h_{\nu}\left(\frac{a \tau +b}{c\tau +d}\right)=(c\tau+d)^{w-1/2}\, h_{\mu}(\tau) \, M(\gamma)_{\mu \nu}\, ,
 \end{equation} 
 and admits a Fourier expansion 
 \be
 h_{\nu} (\tau) = \sum_{\D} C_{\nu}(\D) \, q ^{\D/4k} \, , 
 \ee
where the sum is over the discriminants {$\D = 4kn -\nu^{2}$} for nonnegative integers $n$. The terms in this expansion for which $\D$ is negative are called the polar terms. The Rademacher expansion  for the coefficients  $C_{\nu}(\D)$  is given \cite{Dijkgraaf:2000fq, Manschot:2007ha} by
  \bea\label{radi}
 C_{\nu} (\D) &= & i^{-w + \half} \sum_{c=1}^\infty \left(\frac{c}{2\pi}\right)^{w-\frac{5}{2}} \, \sum_{\wt \D < 0} \, C_{\mu}(\wt \D)   Kl(n, m, c;  \nu, \mu) \  \left\lvert\frac{ \wt \D}{4 k}\right\rvert^{\frac{3}{2}-w} I_{\frac{3}{2}-w}
 \biggl[ \frac{\pi}{c} \sqrt{| \wt \D|  \D}
\biggr] 
\eea
which is determined essentially by the polar terms.

A weak Jacobi form of particular interest to us (see \S\ref{Micro}) is 
\bea\label{ourmicro}
F(\tau, z) &=&  \frac{\vth_1(\t, z)^{2}}{\eta(\t)^{6}} \, ,
\eea  
where $\vartheta_{1}$ is the odd Jacobi theta  function and  $\eta$ is the Dedekind function. 
It has index $k=1$ and weight $w= -2$  with the theta  expansion
\be\label{jacobi-theta2}  
F(\t,z)  = h_0(\t) \, \vartheta_{1,0}(\t, z)\, +  h_1(\t) \, \vartheta_{1,1}(\t, z)\,    . 
\ee
which gives a 2-component  vector-valued modular form $\{ h_{\mu} \}$  with weight $w-1/2= -5/2$ and $\mu = 0, 1$.  
Note that $\mu$ is determined completely by $\D$ as $\mu = \D \, \mod \, 2$, and there is  only a single polar term $(\mu=1, \D=-1)$. As a result,  the Rademacher expansion simplifies to:
 \be\label{rademsp} 
 C(\D) =   2{\pi} \, \big( \frac{\pi}{2} \big)^{7/2}\, i^{5/2} \, \sum_{c=1}^\infty 
  c^{-9/2} \, Kl_{c}(\D) \;  I_{7/2} \big(\frac{\pi \sqrt{\D}}{c} \big)  \, , 
\ee
where we have defined the Kloosterman sum $Kl_{c} (D) $  by
\bea
\label{kloos}
Kl_{c}(\D) := \sum_{\genfrac{}{}{0pt}{}{-c \leq d< 0;}{(d,c)=1}} e^{2\pi i \frac{d}{c} (\D/4)} \; M^{-1}(\gamma_{c,d})_{\nu 1} \; 
e^{2\pi i \frac{a}{c} (-1/4)} \qquad  \,   \,    \qquad  
\eea
with $\nu = \D \, \mod \, 2$ and $ad = 1 \, \mod \, c$. 

\subsection{Kloosterman Sums and Quantum Degeneracies of Black Holes\label{Micro}}

 Consider Type-IIA compactified on $X \times T^{2}$ where $X$ is either $T^{4}$ or $K3$. We use $\CN=2$ truncation of these theories to label the charges \cite{LopesCardoso:1999cv, Dabholkar:2010uh}.
 Consider a state with charge $( q_{0}, q_{1}, q_{a} | p^{0}, p^{1}, p^{a} )$.  Here $q_{0}$ is the number of D0-branes, 
 $q_{1}$ is the number of D2-branes wrapping $T^{2}$, and $q_{a}$ is the number of D2-branes wrapping the two cycles $\Sigma^{a}$ in $X$.  Similarly, $p^{0}$ is the number of D6-branes wrapping  $X \times T^{2}$, 
 $p^{1}$ is the number of D4-branes wrapping $X$,  and $p^{a}$ is the number of D4-branes wrapping 4-cycles $\tilde\Sigma_{a}\times T^{2}$ where $\tilde \Sigma^{a}$ are the dual 2-cycles in X  with $\tilde \Sigma_{a} \cap\Sigma^{b} = \delta_{a}^{b}$.  We see that  $a= 2, \ldots n_{v}$ where $n_{v}-1$ is the number of 2-cycles on $X$; thus $n_{v}=7$ for $T^{4}$ and $n_{v} =23
$ for $K3$. The  electric and magnetic charge vectors  $Q$ and $P$ are given by
 \bea
Q  =  ( q_{0}, -p^{1}; q_{a}) \, ,  \qquad
P  = (q_{1}, p^{0}; p^{a} ) \, , \qquad a, b = 2, \ldots n_{v} \, . 
\eea
 Let $C_{ab}$ be  the intersection matrix for the  2-cycles of $X$. 
Then the T-duality invariants for these charge vectors  (in the heterotic-like  frame  by analogy with the $\CN=4$ theory)
are defined by
\begin{equation}
 Q^2=-2q_0p^1+C_{ab}q^aq^b,\qquad P^2=C_{ab}p^ap^b,\qquad Q.P=-q^1p^1+C_{ab}q^ap^b \, . 
\end{equation}

\begin{myenumerate}

\item\textbf{One-half BPS states}:
When $X = K3$ we obtain a theory with $\CN=4$ supersymmetry in four dimensions.
We consider half-BPS states in this theory. 
 The simplest realization of such a state has only two nonzero charges $q_{0}, p^{1}$. 
The degeneracies $d(N)$ depend only on the invariant\footnote{In our conventions $q_{0}$ is negative and $p^{1}$ is positive.}
\be
N = Q^{2}/2 = -q_{0} p^{1}
\ee
 and are given \cite{Dabholkar:1989jt,Dabholkar:1990yf} in terms of the Fourier coefficients $c(N)$ of  $F(\tau)$ in~\eqref{half-BPS} :
 \be
 d(N) = c(N) \, . 
 \ee
 
 \item\textbf{One-eighth BPS states}:
When $X = T^{4}$ we obtain a theory with $\CN=8$ supersymmetry in four dimensions.
We consider one-eighth BPS states in this theory. The simplest realization of such a state has only five nonzero charges  $q_{0}, q_{1}, p^{1}, p^{2}, p^{3}$. 
The degeneracies $d(\Delta) $
depend only on the quartic U-duality invariant  of charges
\be
\D = Q^2P^2-(Q.P)^2
\ee
and are given \cite{Maldacena:1999bp, Shih:2005qf, Sen:2008ta} in terms of the Fourier coefficients $C(D)$  of  $F(\tau, z)$ in  \eqref{ourmicro}:
\be
d(\Delta) = (-1)^{\D +1} C(\D) \, . 
\ee

 \end{myenumerate}
 
 The near horizon geometry of the  black holes corresponding to these BPS states  has an $AdS_{2}$ factor and the quantum entropy of these black holes is defined via a functional integral of supergravity in the framework of  $AdS_{2}/CFT_{1}$ holography. In the next section we turn to the evaluation of this functional integral.  The Rademacher expansions \eqref{kloos1} and \eqref{kloos} of the Fourier coefficients are particularly suited to compare  the microscopic degeneracies with the quantum entropy obtained  from this  supergravity functional integral in $AdS_{2}$. 
The  terms with $c>1$ are the nonperturbative corrections which in particular involve the Kloosterman sums and our main goal is to see how the path integral can capture this information.

 The structure of the Kloosterman sums  \eqref{kloos1} and \eqref{kloos} is  substantially different in the two cases. Remarkably, the Chern-Simons terms in the $\CN=8$  and $\CN=4$ theories are sufficiently different to correctly reproduce these two very different Kloosterman sums.  The duality invariance of the degeneracy is not immediately obvious from the functional integral  because individual terms  depend on various  charges and not only on the duality invariants.  Fortunately,  properties of Kloosterman sums  make it possible to assemble the final answer into a  duality invariant form  as we discuss briefly in \S\ref{Duality}.
 
 \section{Localization and Nonperturbative Contributions \label{Local}}

The supersymmetry of the formal functional integral makes it possible to apply localization techniques \cite{Dabholkar:2010uh, Banerjee:2009af} to evaluate it. For this purpose we use the $\CN=2$ truncation of the massless sector and consider only F-type chiral terms of the action for the  vector multiplets coupled to the gravity multiplet. It is convenient to use the 
off-shell formalism for  $\mathcal{N}=2$ supergravity  developed in \cite{deWit:1979ug, deWit:1984px, deWit:1980tn} 
in which the vielbein and its superpartners reside in the Weyl multiplet. In addition, we  consider  $n_{v} +1$ vector multiplets with the field content
\be\label{Vectorfields}
{\bf X}^{I} = \left( X^{I}, \O_{i}^{I}, A_{\mu}^{I}, Y^{I}_{ij}  \right) \, , \quad I = 0,  \ldots, n_{v}\, .
\ee
For each $I$,  the multiplet contains eight bosonic and eight fermionic degrees of freedom: $X^{I}$ is a complex scalar, the gaugini $\O^{I}_{i}$ are an $SU(2)$ 
doublet of chiral fermions, $A^{I}_{\mu}$ is a vector field, and $Y^{I}_{ij}$ are an $SU(2)$ triplet of 
auxiliary scalars. 

The $\CN=2$ truncation of the $\CN=8$ theory has $n_{v} =7$\footnote{We use the truncation described in \cite{Dabholkar:2011ec}. We first reduce $N=8$  to $N=4$ by dropping $4$ gravitini of $N=4$.  This amounts to keeping only the NS-NS  charges of the theory that belong to the $12$ of the T-duality group $SO(6, 6)$. We then reduce further to $N=2$ by dropping two gravitini of $N=2$ which leaves behind $8$ vector fields or $n_{v}=7$.}. 
The chiral F-type  action for these fields is described by  the prepotential
\begin{equation} \label{N8prep}
 F(X)=-\frac{1}{2}\frac{X^1}{X^0}\sum_{a,b=2}^{7} C_{ab}X^aX^b \, .
\end{equation}
This prepotential describes the effective action obtained from the dimensional reduction of the two-derivative classical action. Because of $\CN=8$ supersymmetry, there are no quantum corrections of these F-type terms.

The $\CN=2$ truncation of the $\CN=4$ theory has $n_{v} = 23$. 
The  generalized prepotential \cite{LopesCardoso:1999ur, Mohaupt:2000mj} in this case  takes the form
\begin{equation}\label{prepotential}
 F(X,\wh{A})=-\frac{1}{2}\frac{X^1}{X^0}\sum_{a,b=2}^{23}C_{ab}X^aX^b +\frac{\wh{A}}{64}\frac{X^1}{X^0} \, .
\end{equation}
The first term is the classical  prepotential as above but now there is a one-loop correction captured by the second term. There are additional contributions at one loop level  from world-sheet instantons  but these can be ignored as they vanish in the background of small black holes because of additional fermionic  zero modes.

Near the  horizon, the  metric is $AdS_{2} \times S^{2}$ and  the values of the scalar fields are given by 
\be\label{attract}
X^{I}_{*} = \frac{1}{2}(e_{*}^{I} + i p^{I})\, , \qquad \wh{A}_{*} = 64 \, ,
\ee
where $e_{*}^{I}$ are the attractor value of the electric fields  determined in terms of the charges $(q, p)$. The $AdS_{2}$  functional integral is defined by summing over all field
configurations which asymptote to the these attractor values with the fall-off conditions \cite{Sen:2008yk, Sen:2008vm, Castro:2008ms}:
\bea\label{asympcond}
d s^2 &=& v_{*} \left[ 
\left(r^2+\cO(1)\right) d\theta^2+ \frac{dr^2}{r^2+\cO(1)}  \right]\,  ,\\
X^{I} &= &X^{I}_{*} + \cO(1/r)\ ,\qquad
A^I = -i  \, e_{*}^{I} (r -\cO(1) ) d\theta\ .
\eea
To make the classical variational problem well-defined,   it is necessary to add a boundary term to the action so the equations of motion for the gauge field are satisfied everywhere.
 With this boundary term, the quantum bulk partition can be naturally interpreted as  an expectation value of a Wilson line inserted at the boundary
\be\label{qef}
W (q, p) = \left\langle \exp[ \frac{i}{2} \, q_I \oint_{\theta}  A^I]  \right\rangle_{\rm{AdS}_2}^{finite}\ .
\ee 
The superscript refers to a finite piece obtained by a holographic renormalization procedure to remove a divergence coming from the infinite volume of the $AdS_{2}$~\cite{Sen:2008yk}.

\subsection{Localization in Supergravity}\label{localization sec}

It was shown in  \cite{Dabholkar:2010uh} from the analysis of the vector multiplets that the functional integral \eqref{qef} localizes on a 
solution  parametrized by $(n_{v} +1)$ real parameters $\{C^{I}\}, \ I = 0, \ldots, n_{v},
$ and is given by the  field configurations
\begin{eqnarray} \label{HEKEsol}
 X^{I}  =  X^{I}_{*} +  \frac{C^{I}}{r} \ , \qquad  \bar X^{I}  =  \bar X^{I}_{*} +  \frac{C^{I}}{r} \ , \qquad Y^{1I}_{1} = - Y^{2I}_{2} =  \frac{2C^{I}}{r^{2}} \,\,  , \end{eqnarray}
with other fields fixed to their attractor values. Note that the auxiliary fields $Y^{I}_{ij}$ have a nontrivial profile indicating that the localizing saddle point is far away from the physical minimum. 
It was shown later in \cite{Gupta:2012cy} that this is the most general solution even after including the Weyl multiplet\footnote{This result was derived assuming that the auxiliary $SU(2)$ gauge fields in the Weyl multiplet
vanish. The analysis suggested that these fields can be eliminated 
upon including hypermultiplets in the off-shell theory.}.  The real parameters 
$\{C^{I}\} $ are the collective coordinates of the localizing instantons.  The functional integral 
of supergravity thus reduces to  a finite dimensional ordinary bosonic integral over  $\{C^{I}\}$ with integrand the exponential of 
a \emph{renormalized action}, thus leading to an  enormous simplification \cite{Dabholkar:2010uh}.  Using new variables 
\be\label{ephi}
\phi^I := e_{*}^I+2 C^I \ ,
\ee
with $ e_{*}^I$ defined by \eqref{attract}, we can express the renormalized action as
\begin{eqnarray} \label{Sren}
 \mathcal{S}_{ren}(\phi, q, p) =  - \pi  q_I   \phi^I + \mathcal{F}(\phi, p)\, ,
\end{eqnarray}
with
\begin{equation} \label{freeenergy2}
\mathcal{F}(\phi, p) = - 2\pi i \left[ F\Big(\frac{\phi^I+ip^I}{2} \Big) -
 \bar{F} \Big(\frac{\phi^I- ip^I}{2} \Big) \right] \, ,
 \end{equation}
which has the same form conjectured in \cite{Ooguri:2004zv}. Note that, in this expression, the prepotential is evaluated precisely for values of the scalar fields at the \textit{origin} of~$AdS_{2}$ for the localizing solution and not at the boundary of~$AdS_{2}$.  At the boundary, the fields remain pinned to their attractor values  and in particular  the electric field remains fixed  as required by the microcanonical boundary conditions of the functional integral. The collective coordinates  $\phi^{I}$  in \eqref{ephi} fluctuate in the functional integral because $C^{I}$, the parameters of the off-shell solutions, take values over the  real line. 
 
For the one-eighth BPS black holes in the $\CN=8$ theory, the horizon area is large. Moreover,   the classical prepotential \eqref{N8prep} is quantum-exact. The effective action involves only two-derivatives and the measure on the localizing submanifold  is determined by this action\footnote{There are subtleties in gauge fixing to Poincar\'e gravity which can affect the measure. See \S\ref{Assess}.}. Evaluating the integral over the 
parameters~$\phi^{I}$ with $n_{v}=7 $,  we obtain the Bessel function $I_{7/2} \big( \pi \sqrt{\D} \big)$
which is precisely the $c=1$ term of the corresponding microscopic degeneracy \eqref{rademsp} 
\cite{Dabholkar:2011ec}.  

\subsection{An $SL(2,\mathbb{Z})$ family of Localizing  Solutions \label{flysols}}

The four-dimensional  $AdS_2\times S^2$ geometry  in the Type-IIA frame corresponds to a five-dimensional $AdS_2 \times S^1\times S^2$ geometry  in the M-theory frame. 
 It can be shown  \cite{Gomes:2013cca} using localization that the entropy functional in five dimensions reduces exactly to four dimensions\footnote{The results of   \cite{Gomes:2013cca} are in the context of $AdS_2\times S^3$  but  extend easily  to $AdS_2\times S^1\times S^2$.} to give the quantum entropy function (\ref{Sren}).
We reduce the  five-dimensional theory on the $S^2$ to obtain\footnote{The reduction on the $S^{2}$  requires care because of Dirac string singularities \cite{deWit:2009de}.} a three-dimensional theory. This three-dimensional point of view will be particularly useful especially when we discuss the contributions from the Chern-Simons terms.  

{We now describe a family of supersymmetric orbifolds which all  satisfy the $AdS_{2}$ boundary conditions and admit  localizing solutions. Contributions from these additional localizing saddle points precisely give rise to the nonperturbative corrections to the quantum entropy.}

{Let $J$ be a generator of rotations of the round sphere $S^{2}$ and $L$  be the generator of rotations of Euclidean  $AdS_{2}$ which is a disk with Poincar\'e metric.  Consider a $Z_{c}$ action generated by $\exp{\frac{2\pi i}{c}(L-J)}$. This   preservers supersymmetry.  One can accompany this twist by a shift along the coordinate of the $S^{1}$ by $2\pi d/c$ for $d$ relatively prime to $c$ to obtain a freely acting orbifold\footnote{It may appear that one can have  more general shifts along the $T^{6}$. However, we are considering an orbifold not in flat space but in the presence of the black string from the 5-dimensional  point of view.  If the magnetic  charges of the black hole are relatively prime then the identifications along the internal torus will violate the flux quantization and hence are not allowed \cite{Sen:2009gy}.}. This gives a family of supersymmetry preserving smooth orbifolds labeled by two integers $(d, c)$.  The geometry of the $AdS_{2} \times S^{1}$ part of the orbifold is crucially important for our purposes which we now discuss. The action on the $S^{2}$ contributes to a phase which we discuss later in \S{\ref{Phases}}. }

Before taking any orbifold, the metric for  $AdS_2 \times S^1$  is 
\begin{equation}\label{metric ads2xs1}
 ds^2= (\tilde r^2-1)d \tilde \theta^2+\frac{d\tilde r^2}{\tilde r^2-1}+R^2\left(d\tilde y - \frac{i}{R}(\tilde r-1)d \tilde \theta\right)^2 \, 
\end{equation} 
with standard identifications
\begin{equation}\label{standard ident}
 (\tilde \theta, \tilde y)\sim \left(\tilde \theta+ {2\pi }, \tilde y \right)\sim (\tilde \theta, \tilde y+2\pi)\, 
\end{equation}  
where $R$ is the attractor value of the M-theory circle near the black hole horizon.
We see that  the M-circle is  fibered over $AdS_{2}$ because  the Kaluza-Klein gauge field couples to the M-momentum which in the Type-IIA frame is  the D0-brane charge $q_{0}$.

The orbifolds must preserve the $AdS_{2}$ boundary conditions \eqref{asympcond} of the functional integral. Such orbifolds are labeled by two integers $(c, d)$. For every  positive integer $c$, the generator for the orbifold symmetry is a supersymmetry preserving $\mathbb{Z}_c$ twist on $AdS_2\times S^2$  together with a shift of $d$ units along the  M-circle $S^1$. Hence $d$ is defined only mod $c$.  If $c$ and $d$ are relatively prime then the orbifold symmetry is freely acting without fixed points and one obtains a smooth manifold $\mathcal{M}(c, d)$. In the  coordinates used in~\eqref{metric ads2xs1}, the metric for the orbifold $\mathcal{M}(c, d)$ has the same form as for the unorbifolded manifold $\mathcal{M}(1, 0)$ but the  identifications are different:
\begin{equation}\label{orbifold ident}
 (\tilde \theta, \tilde y)\sim \left(\tilde \theta+\frac{2\pi }{c}\, , \, \tilde y + \frac{2\pi d}{c}\right)\sim (\tilde \theta, \tilde y+2\pi)\, .
\end{equation}  
  
The family of distinct extremal solutions given by \eqref{metric ads2xs1} with periodicities~\eqref{orbifold ident} all have the same asymptotics as in \eqref{asympcond}. To see this explicitly, it is convenient to use new coordinates 
\be
( \tilde r, \tilde \theta,  \tilde y )  = (c \, r \, , \frac{ \theta}{c}\, ,  {y} +  \frac{d}{c} \theta )
\ee
 so that the identifications of the new coordinates $ \theta$ and $ y$ 
\begin{equation}\label{orbifold ident again}
 ( \theta,  y)\sim \left(  \theta + {2\pi },  y \right)\sim ( \theta,  y+2\pi)\, 
\end{equation}
are the same as  \eqref{standard ident} for  the unorbifolded theory.
The metric now takes the form
\begin{equation}\label{new metric ads2xs1}
 ds^2= ( r^2-\frac{1}{c^{2}})d\theta^2+\frac{d r^2}{ r^2- \frac{1}{c^{2}}}+R^2\left(d y - \frac{i}{R}(r- \frac{1}{c}) d \theta + \frac{d}{c}d  \theta \right)^2  \, .
 \end{equation} 
Thus,  the KK gauge field has  the form
\begin{equation}\label{gauge field shift}
 A=  - \frac{i}{R}(r- \frac{1}{c}) {d  \theta} + \frac{d}{c}d \theta = A_{(0)} +  \frac{d}{c}d  \theta ,
\end{equation}
 where $A_{(0)}$ is the  KK gauge field  when there is no shift along the M-circle.
 It is clear that the  large $r$  behavior  of the metric and the gauge fields is the  same for all $\mathcal{M}(c, d)$. Hence all these orbifold geometries will contribute to the  functional integral defined with the boundary conditions \eqref{asympcond}.  
 The integers~$(c,d)$ are  naturally identified \cite{Banerjee:2008ky, Murthy:2009dq} with the integers appearing in the microscopic degeneracy \eqref{classical Kl} and \eqref{gener Kloosterman}. 

It is   convenient to use a particular representation of these orbifolds \cite{Murthy:2009dq} obtained as a limit of orbifolds of global $AdS_{3}$. 
There is a well-known $SL(2, \mathbb{Z})$ family of configurations labelled by $(c,d)$, that all have the same $AdS_{3}$ asymptotics \cite{Maldacena:1998bw}. 
These configurations are crucial in the ``Farey Tail''  
interpretation of $AdS_{3}/CFT_{2}$  \cite{Dijkgraaf:2000fq, Manschot:2007ha, deBoer:2006vg}.  
We now briefly review this construction. 

A cover of global  $AdS_3$  has the metric
\begin{equation}\label{global AdS3}
 ds^2=- \cosh^{2} \eta \, dt^2+\sinh^{2} \eta \, d\chi^2+d\eta^2
\end{equation}and $-\infty<t<\infty$, $0\leq\eta<\infty$ and $0\leq\chi\leq 2\pi$. The Euclidean section is obtained by the usual Wick rotation $t\rightarrow i t_E$ with  $-\infty<t_E<\infty$. Thermal $AdS_3$ is obtained by further identifications on the cylinder 
\begin{equation}\label{thermal identification}
 \chi+it_E\sim \chi+i t_E+2\pi\sim \chi+i t_E+2\pi\tau
\end{equation}with $\tau$ a complex number in the upper half plane.
Topologically this is a solid torus $D^2\times S^1$ where $D^2$ is an open disc parametrized by $(r,\chi)$. The boundary is a torus with  complex structure parameter $\tau$ and holomorphic 1-form $\omega : =  d\chi + i dt_{E}$.  The  cycle $C_{2}$  parametrized by $\chi$ is  contractible and the noncontractible cycle $C_{1}$ is  such \footnote{We choose orientation such that  $C_{2} \cap C_{1} =1$.}  that   $\oint_{C_{2}}\omega= 2\pi $ and  $\oint_{C_{1}} \omega= 2\pi \tau$.

The $SL(2, \mathbb{Z})$ family of solutions is obtained from this basic solution by choosing a different homology class for  the boundary cycle that becomes contractible  in the full geometry \cite{Maldacena:1998bw,Maloney:2007ud}.
To do so , consider a `Dehn twist'  on the boundary torus which relabels the cycles
\begin{eqnarray}\label{Dehn fill}
 &\left(\begin{array}{c}
   C_{n}\\
   C_{c}
   \end{array}\right)=\left(\begin{array}{cc}
                             a & b\\
			     c & d
                            \end{array}\right) \left(\begin{array}{c}
							 C_{1}\\
							C_{2}
                                                     \end{array}\right)\,  \qquad \text{for} \qquad 
                                                     \left(\begin{array}{cc}a & b \\c & d\end{array}\right) \in SL(2, \mathbb{Z})  \, ,
\end{eqnarray}
and takes the complex structure parameter to 
\be
                                                      \tau'=\frac{\oint_{C_{n}}\omega}{\oint_{C_{c}}\omega}=\frac{a\tau+b}{c\tau+d}\, .
\ee
One would now like to fill in a smooth manifold inside  this `Dehn-twisted'  boundary torus. Topologically, this  operation  is called  `Dehn-filling' and in fact it is possible to write a smooth hyperbolic metric on this manifold~\cite{Maldacena:1998bw}.
In this Dehn-filled geometry now  the cycle $C_{c}$  becomes contractible instead of the cycle $C_{2}$ of the original thermal $AdS_{3}$. In particular, 
for $(c,d)=(1, 0)$ the contractible cycle is now   $C_{1}$ which is  along the time direction   and hence  it  represents a Euclidean continuation of nonextremal  BTZ black hole~\cite{Banados:1992wn}.  In the coordinates \eqref{orbifold ident again} the cycle  $C_{1}$ is  parametrized by $-\theta$ and $C_{2}$ is parametrized by $y$.

The  geometries $\mathcal{M}(c, d)$ of interest are  orbifolds of supersymmetric \textit{extremal} BTZ black hole. The extremal limit of the geometries above is obtained by taking $\bar{\tau}\rightarrow \infty$, keeping $\tau$ fixed~\cite{Strominger:1998yg}. This corresponds heuristically to taking the right-moving temperature in the boundary thermal CFT to zero so that the right-moving super symmetries are preserved. This is possible in the Lorentzian section in which $\tau$ and $\bar{\tau}$ are no longer complex conjugates.  
In contrast to the $AdS_{3}$ Farey tail, the charge of the BTZ black hole  and hence $R$ is fixed while the parameter $\tau$ of the boundary $AdS_{3}$  varies.
To obtain the $\mathcal{M}(c, d)$  geometry \eqref{new metric ads2xs1} with  given parameters ~$R, c, d$, one must choose \cite{Murthy:2009dq}  the complex structure $\tau$  of the~ thermal $AdS_{3}$ such that 
\be\label{R}
 i R = \frac{1}{c\tau +d}\, .
 \ee 

One can now examine the  localization solutions in the orbiold geometry.
Because the space is still locally  $AdS_2 \times S^{1}\times S^2 $, the localization equations are not changed and therefore we obtain the same solutions (\ref{HEKEsol}) for the orbifolds. However, the renormalized bulk action gets reduced by a factor of $1/c$.  As a result the Bessel functions in the Rademacher expansions \eqref{rademsp} and \eqref{kloos1} with an argument reduced by $c$ are correctly reproduced for all $c$. 
The question that we  now would like to address is the origin of the Kloosterman sums. 

\subsection{Boundary Terms for the Chern-Simons Action}\label{sec CS bnd terms}

The analysis in \S\ref{localization sec} is local  and is insensitive to the global properties of the orbifolds.  However, the Wilson lines and  the Chern-Simons terms for  various gauge fields in the action  are sensitive to global properties of the space $\mathcal{M}(c, d)$ that depend on $c$ and $d$.  These lead to additional contribution to the renormalized action and additional localizing saddle points specified by the holonomies of flat connections of these gauge fields.  These lead to additional phases. 

{The gauge group in the bulk supergravity has many factors. There are several $U(1)$ factors corresponding to the $n_{v}$ vector multiplets. In addition there is an  $SU(2)_{L} \times SU(2)_{R}$ factor corresponding to the R-symmetry (after an appropriate dimensional uplift).  To discuss the contributions from the higher derivative terms, it is also useful to consider the Chern-Simons formulation of gravity in $AdS_{3}$ in \S\ref{Sec grav CS} with an $SL(2)_{L} \times SL(2)_{R}$ gauge group.
Remarkably various phases in  \eqref{classical Kl} and \eqref{gener Kloosterman} arise from intricate combinations of the boundary terms and Chern-Simons terms of these various gauge groups in the bulk supergravity. } The contribution to the phases from the $U(1)$ factors, the $SU(2)_{R}$ factor, and the $SL(2)_{L} \times SL(2)_{R}$ factor is discussed in \S\ref{Phases}. The contribution from the $SU(2)_{L}$ is discussed in \S\ref{Multiplier}.
In this subsection we discuss  the general structure of the Chern-Simons action  and the associated boundary terms for these gauge groups of interest.

Consider the Chern-Simons action for a gauge field $A$  on the manifold $\mathcal{M}(c, d)$
\beq\label{CS}
I(A) = \int_{\mathcal{M}}\text{Tr}\left(A\wedge dA+\frac{2}{3}A^3\right).
\eeq
For abelian gauge fields the trace is trivial and the cubic term is absent. For nonabelian $SL(2)$ and $SU(2)$ gauge fields  the trace is in the fundamental representation.

Asymptotically all $\mathcal{M}(c, d)$ geometries are $AdS_{2} \times S^{1}$ and  have a torus boundary.   One must introduce appropriate boundary terms to obtain a well defined variational problem. We describe these boundary conditions  in the  reference geometry  $\mathcal{M}(1,0)$.  
In the basis of cycles $(C_{1}, C_{2})$ in (\ref{Dehn fill}),  the components of the gauge field at the boundary are $A_{1}$ or $A_{2}$. Because the Chern-Simons action is first order,  we can fix only one component at the boundary.  
To determine  the correct boundary conditions we relate the 3d theory to the 2d theory in $AdS_{2}$ where the boundary conditions are specified by \eqref{asympcond}.
The geometry $\mathcal{M}(1,0)$ is  the unorbifolded extremal BTZ black hole obtained from thermal $AdS_{3}$  by a Dehn-twist by the element $S=\left(\begin{smallmatrix} 0&-1\\ 1&0 \end{smallmatrix}\right)$ of $SL(2, \mathbb{Z})$. Hence,  the contractible cycle is now $C_{1}$  and corresponds to the boundary of $AdS_{2}$ with coordinate $\theta$ in \eqref{new metric ads2xs1} whereas the noncontractible cycle is $-C_{2}$.   

Demanding invariance under all eight supercharges of the five dimensional supersymmetry implies that the three dimensional  gauge fields, that is, the components ``orthogonal'' to the sphere, 
are flat \cite{deWit:2009de}\footnote{{ After localization the configuration needs to be invariant under one supercharge.  This allows for  a non-flat profile for the gauge field, as we discuss later.}}. The component  of the gauge field along $C_{2}$ gives the scalar field in $AdS_{2}$. Hence, the  three-dimensional gauge field takes the form
\beq
A^I_{\mathcal{M}(1,0)}=-\phi_{*}^I R\, dy \, , \qquad I = 1,  \ldots, n_{v}\label{bnd cond flat} \, .
\eeq 
For  Kaluza-Klein reduction we reexpress this field in terms of gauge invariant one-forms:
\beq
A^I_{\mathcal{M}(1,0)}=-\phi_{*}^I R\left(dy - i \frac{1}{R}(r-1)d\theta \right) - i\phi_{*}^I(r-1)d \theta
\eeq 
and  read off the two dimensional field $A_{(2)}^{I}= - i\phi_{*}^I(r-1)d \theta$. 
This is consistent with  the normalization  used in \cite{Dabholkar:2010uh} with $A_{(2)}^{I}=-ie_{*}^{I}(r-1)d\theta$ because $e^{I}_{* }= \phi_{*}^I$. 

In $AdS_2$, the constant mode of the scalar field (coming from the reduction of the 3d gauge field on the non-contractible KK circle) is nonnormalizable and hence is  held fixed at the boundary to its attractor value. On the other hand, the constant mode of the 2d gauge field which is the chemical potential is normalizable and hence is allowed to fluctuate \cite{Sen:2008yk}. For the 3d gauge field this implies  that in the quantum problem its component along the cycle $C_{2}$ is held fixed to set the boundary condition of the functional integral and  the component   along the cycle $C_{1}$ is allowed to fluctuate:
\beq \label{bnd cnd U(1)}
\oint_{C_{2}} A^I=  -2\pi \phi_{*}^IR \, ,\qquad \oint_{C_{1}} A^{I}=\text{not fixed} \, 
\eeq
using the fact that  $y$ is the coordinate of $C_{2}$. In the classical problem, the component  $\oint_{C_{1}} A^{I}$ gets determined by physical requirements on the gauge fields in the interior which are  different for different gauge fields. From this 2d analysis  we conclude following \cite{Elitzur:1989nr} that the correct boundary terms for the 3d theory are given by
\beq \label{CS bnd term micro}
I_{b}(A) = \int_{T^2}\text{Tr}A_1A_2d^2x \, 
\eeq 
Note that  $A_1$ instead of $A_{2}$ were  fixed, then the boundary term would be 
$
-\int_{T^2}\text{Tr}A_1A_2d^2x \, .
$

The same reasoning applies to nonabelian gauge fields. For example, for the $SU(2)_{R}$ gauge field, the Wilson line along the boundary cycle $C_{2}$  of $AdS_{2}$ is the angular momentum $J_R$.  The $AdS_{2}$ boundary conditions imply that $J_R=0$ so that index equals degeneracy \cite{Sen:2009vz}
\beq\label{deg vs ind}
\text{Tr}(-1)^{J_R}=\text{Tr}(1).
\eeq In other words, we fix the Wilson line
\beq \label{ARWilson}
\oint_{C_{2}} A_R=0 \, .
\eeq 
 
The phases that assemble into the Kloosterman sum arise essentially from the boundary Wilson lines and the Chern-Simons action evaluated for the \textit{flat} connections of various gauge fields. We now explain the general philosophy behind these   computations. 
For this purpose, it is useful  to parametrize the Wilson lines along the basis cycles $C_{1}$ and $C_{2}$ by\be\label{dg}
\oint_{C_{1}}A =  2\pi i \gamma \frac{\sigma^3}{2}  \, \qquad \oint_{C_{2}}A =  2\pi i \delta \frac{\sigma^3}{2}\, .
\ee
For a given manifold  $\mathcal{M}(c, d)$ we will need to  know also  the Wilson lines  around the contractible cycle $C_{c}$ and noncontractible cycle $C_{n}$. 
We parametrize these  by two real numbers $\a$ and $\b$ as
\be\label{ab}
\oint_{C_{c}}A =  2\pi i \a \frac{\sigma^3}{2}  \, \qquad \oint_{C_{n}}A =  2\pi i \b \frac{\sigma^3}{2}\, .
\ee
It follows from  \eqref{Dehn fill} that 
\be\label{reln}
\a = c\gamma + d \delta \, , \qquad \b = a \gamma + b\delta \, .
\ee 

The boundary action \eqref{CS bnd term micro} can be readily evaluated given $\gamma$ and $\delta$. Evaluation of the bulk action \eqref{CS} is more nontrivial but using a result \eqref{CS action theorem} from the work of  \cite{kirk1990}, it can be computed easily knowing  $\alpha$ and $\beta$.  Our problem thus reduces to computing  $\gamma$ and $\delta$ because $\a$ and $\b$ are   determined using  \eqref{reln}.
In the physical problem at hand, $\delta$ is specified by the boundary condition \eqref{bnd cnd U(1)} but  $\gamma$ is not specified directly. It has to be determined from physical considerations as we discuss in the next section.

\section{Phases from Wilson Lines and  Chern-Simons Terms \label{Phases}}

In this section we will  evaluate the phases in the Kloosterman sums by two different methods. In  \S\ref{Sec Wl + CS phase} and \S\ref{CS phase} we use the  metric formulation of supergravity which admits an \textit{off-shell} completion to evaluate  the phases. The computation is then justified by localization of the path integral on $\mathcal{M}(c, d)$. In \S\ref{Sec grav CS} we use the  \textit{on-shell} Chern-Simons formulation of gravity to give a simpler rederivation of the same result. The on-shell computation is justified heuristically by the fact that   the phases are essentially topological in origin.

\subsection{Phases from Boundary Wilson Lines and Abelian Chern-Simons terms}\label{Sec Wl + CS phase}
  
To understand the possible phase contributions arising from the abelian sector, we need to first work out the different boundary terms  in five dimensions. This construction is similar to the one presented in \cite{Gomes:2013cca}. Supersymmetry requires that the  abelian gauge field  be flat at infinity~\cite{deWit:2009de}. As a result,  variations of the Maxwell terms do not generate any boundary terms\footnote{In the previous version of the paper these were  treated incorrectly.}.  On the other hand, the Chern-Simons terms require a careful treatment, as explained in section \S\ref{sec CS bnd terms}. These terms after a circle reduction to four dimensions lead to the Wilson lines on the $AdS_{2}$ boundary \eqref{qef}. Since the metric gives rise to a Kaluza-Klein gauge field  in two dimensions with a non-zero electric field, normalizable fluctuations\footnote{By normalizable we mean the fluctuations which leave the electric field fixed as required in  $AdS_2$.} of this metric component lead to an additional boundary term.
  
Consider the five dimensional  abelian Chern-Simons action. To simplify the analysis, we reduce it on an ungauged $S^2$ and ``diagonalize`` the Chern-Simons couplings using the intersection matrix $C_{ab}$ introduced in \S\ref{Micro}:
\begin{eqnarray}\label{5d3d}
&&\frac{\pi i}{3(4\pi)^3}\int_{S^2}c_{IJK}A^I\wedge F^J\wedge F^K=\frac{2\pi i}{(4\pi)^2}C_{ab}A^{1}\wedge F^ap^b+\frac{\pi i}{(4\pi)^2}p^1C_{ab}A^a\wedge F^b\nonumber\\
&=&-\frac{\pi i}{p^1(4\pi)^2}P^2A^1\wedge F^1+\frac{\pi i p^1}{(4\pi)^2}C_{ab}\left(F^a+\frac{p^a}{p^1}F^1\right)\wedge \left(A^b+\frac{p^b}{p^1}A^1\right)
\end{eqnarray} 
with $ a, b= 2, \ldots, n_{v}$. This rewriting simplifies the computation of  on-shell values of the  fields.  The boundary terms for the resulting 3d Chern-Simons action are as in \S\ref{sec CS bnd terms}:
\beq\label{abelian CS bnd}
-\frac{\pi i}{p^1(4\pi)^2}P^2\int_{T^{2}}A^1_{1} A^1_{2}+\frac{\pi i p^1}{(4\pi)^2}C_{ab}\int_{T^{2}} \left(A^a_{1}+\frac{p^a}{p^1}A^1_{1}\right)\left(A^b _{2}+\frac{p^b}{p^1}A^1_{2}\right)
\eeq 
As explained before (\ref{bnd cnd U(1)}) the component $A^I_2$ is fixed\footnote{We use $R=-2/\phi^0_{*}$ which follows from  the $4d$-$5d$ dictionary \cite{Gomes:2013cca}. See  \S\ref{Sec grav CS} for a simpler derivation.} to its attractor value $A^{I}_2=2\phi^I_*/\phi^0_*\;$. 

Let us first consider how to obtain the two dimensional Wilson lines from the Chern-Simons terms in the unorbifolded theory. To perform localization as in \cite{Gomes:2013cca} we need to rewrite the Chern-Simons action in terms of two dimensional quantities, that is, we write the three dimensional gauge field as
\begin{equation}
A^{3I}=\chi^I(dy+A_{KK})+A^{2I}
\end{equation}where $A^{2I}$ is the gauge field living on $AdS_2$. In rewriting this way we generate non gauge-invariant terms in two dimensions.  Integrating by parts solves the problem but gives additional boundary terms via total derivatives. For instance in the simple example of only one gauge field we obtain
\begin{equation}\label{CS decomposition}
\int_{\mathcal{M}(1,0)} A\wedge F=\int_{\mathcal{M}(1,0)} \chi^2 dy\wedge F_{KK}+2\chi dy\wedge F_2-d(\chi A_2\wedge dy).
\end{equation}
Integrating the total derivative gives an additional boundary contribution. To compute the boundary terms we take the off-shell solution allowed by localization. The two dimensional electric fields are fixed to their on-shell values but the scalars $\chi^I$ have a non-trivial profile \cite{Dabholkar:2010uh,Gomes:2013cca}, that is,
\begin{equation}
 \chi^I=2\frac{\phi^I_*+C^I/r}{\phi^0_*+C^0/r}.
\end{equation}
with $C^I,C^0$ constants. The total boundary term in two dimensions coming from the treatment of the Chern-Simons terms thus has two parts: 
\begin{eqnarray}
CS_{\text{bnd}}+\text{total deriv.}=\textbf{Finite}\Big[-\frac{\pi P^2}{8p^1}(\chi^1)^2\phi_*^0(r-1)+\frac{\pi p^1}{8}C_{ab}\tilde{\chi}^a\tilde{\chi}^b\phi_{*}^0(r-1)\\
+\frac{\pi P^2}{2p^1}\chi^1\phi_*^1(r-1)-\frac{\pi}{2} p^1 C_{ab}\tilde{\chi}^a\tilde{\phi}^b_*(r-1)\Big]
\end{eqnarray}
where we denote the combination $\tilde{\chi}^a=\chi^a+p^a/p^1 \chi^1$ and similarly for $\tilde{\phi}^a=\phi^a+p^a/p^1 \phi^1$, and "finite" stands for the finite piece as $r\rightarrow \infty$. We have to be careful in taking the limit because $\chi$ has subleading terms in $1/r$. We find that the final result is independent of these subleading terms and we obtain
\begin{equation}\label{CS bnd Wilson}
CS_{\text{bnd}}+\text{total deriv.}=-\frac{\pi}{2}q_1\phi^1_*-\frac{\pi}{2}q_a\phi^a_*=\textbf{Finite}\left[ i\frac{q_J}{4}\oint A^{2J}\right]
\end{equation}
where $q^J$, with $J=1\ldots n_v$, are the two dimensional charges. It is interesting to observe that these boundary contributions do not depend on the value of the gauge field along the cycle $C_1$. To appreciate this, note that for very large $r$ we have
\begin{equation}\label{connection at infty}
A^{3I}=\chi_*^Idy+\frac{i}{2}C^Id\theta-\frac{i}{2}\chi_*^I C^0d\theta+{\mathcal{O}(1/r)}
\end{equation}where $C^I$ are the constant modes allowed by localization. We will argue that this is also true in the orbifolded theory so that only the KK Wilson line boundary term gives a non-trivial phase.

We now consider the boundary term for the Kaluza-Klein gauge field. After reducing the five dimensional Ricci scalar down to $AdS_2$ we obtain a Maxwell term
\begin{equation}\label{KK reduction}
\int_{AdS_2\times S^1\times S^2} d^5x\sqrt{g} \mathcal{R}\rightarrow \int_{AdS_2} d^2x\sqrt{g} \left[R\mathcal{R}-\frac{1}{4}R^3F_{KK}^2+R\mathcal{R}_{S^2}\right]
\end{equation}where $\mathcal{R}$ is the Ricci scalar (not to be confused with the fiber radius $R$), and $F_{KK}$ is the Kaluza-Klein gauge field strength. The Maxwell term gives rise to a boundary Wilson line as explained in \cite{Gomes:2013cca,Sen:2008vm} \footnote{There is an overall charge dependent factor multiplying the five dimensional Ricci scalar. This factor is proportional to $C_{IJK}\sigma^I\sigma^J\sigma^K$, where $\sigma^I$ is the five dimensional vector multiplet scalar which has on-shell value $\sigma^I=p^I$, the magnetic charge \cite{deWit:2009de}. The coefficient of the Ricci scalar is proportional to $p^1P^2$ with the correct normalization fixed by the value of the central charge computed in three dimensions.}
\begin{equation}\label{KK wilson line}
i\frac{p^1P^2}{8}R^2\oint A_{KK}=i\frac{\Delta}{2p^1 P^2}\oint A_{KK}.
\end{equation}We have used the on-shell value of $R$ (\ref{on-shell val R}). Putting this boundary term together with (\ref{CS bnd Wilson}) we obtain, at the on-shell level, the two dimensional Wilson lines
on~$AdS_{2}$ as indicated in~(\ref{qef}):
\begin{equation}
\textbf{Finite}\left[i\frac{\Delta}{2p^1 P^2}\oint A_{KK}+i\frac{q_I}{4}\oint A^{2I}\right]=\textbf{Finite}\left[i\frac{q_I}{2}\oint A^I\right]
\end{equation}
where the index on the RHS of the expression now runs in the interval $I=0\ldots n_v$ and we have used the on-shell values of the fields 
\begin{eqnarray}
 &&\phi_{*}^a=-q^a\frac{\phi_{*}^0}{p^1}+\frac{Q.P}{P^2}\frac{\phi_{*}^0}{p^1}p^a \\
 &&\phi_{*}^1=-\frac{Q.P}{P^2}\phi_{*}^0 \,  
 \end{eqnarray}obtained via a saddle point of the entropy function (\ref{Sren}). The bulk gauge-invariant action living in $AdS_2$ is the same as considered originally in \cite{Dabholkar:2010uh} and therefore the action on the localization locus including the boundary terms gives precisely the Bessel function as derived in~\cite{Dabholkar:2011ec}.

Let us now consider  the orbifolded theory. From a ten dimensional point of view, the geometry is locally $AdS_2\times S^2\times T^6$. Instead of taking the ten dimensional geometric point of view, we need to consider the reduced five dimensional theory for which we have an off-shell representation. Orbifold action of shifts  can be  equivalently interpreted as turning on a constant mode  for the abelian fields living in five dimensions. Localization fixes the two dimensional gauge fields to their on-shell values \cite{Gomes:2013cca,Dabholkar:2010uh} and since these equations only allow for smooth field configurations living on $AdS_2$ the 3-dimensional gauge field is completely determined to be\footnote{In the previous version of the paper the two dimensional gauge fields had delta function singularities at the origin which however is forbidden by localization. }
\begin{equation}
A^{3I}= 2\frac{\phi^I}{\phi^0} \left(dy-\frac{i}{R}(r-1)d\tilde{\theta}\right)-i\phi_{*}^I(r-1)d\tilde\theta,
\end{equation}where both $\phi^I$ and $\phi^0$ have the localization profile and $\tilde{\theta}$ has periodicity $2\pi/c$. Note that $A^{3I}$ asymptotes for large $r$ to (\ref{connection at infty}) and therefore we see that, from a three dimensional point of view, we are in fact integrating over the constant mode. The computation is similar to the one in the unorbifolded theory except that now the volume gets divided by a factor of $c$. 

On the other hand, the boundary Wilson line  (\ref{KK wilson line}) for the KK gauge field gives a non-trivial phase via the shift (\ref{orbifold ident})
\begin{equation}\label{WL phase}
i\frac{\Delta}{2p^1 P^2}\oint A_{KK}\rightarrow i\pi\frac{\Delta}{p^1 P^2}\frac{d}{c}.
\end{equation}Note that in this case there is no bulk singularity. The orbifold is freely acting and as a consequence the curvature is smooth everywhere. The KK reduction is always of the form (\ref{KK reduction}) with the field strength $F_{KK}$ smooth on $AdS_2$ with the volume reduced by a factor of $c$. We will present a different derivation of this result in  \S\ref{Sec grav CS} where we show that this phase comes precisely from the boundary terms in the  $SL(2,\mathbb{R})$ Chern-Simons formulation.

For the black hole under consideration we have $P^2=2$ and $p^1=1$ so that the $SU(2)_R$ level becomes $k_{R} = 1$. At the on-shell level the three dimensional gauge field is given simply by
$
 A^{3I}= 2\phi^I_*/\phi^0_* dy
$ and the holonomy along the contractible cycle $cC_1+dC_2$ becomes trivial as required
\begin{eqnarray}
 &&\exp{[i\oint_{cC_1+dC_2} A^a]}= \exp{\left[4\pi i d (-q^a+\frac{Q.P}{2}p^a)\right]}=1\\
&&\exp{[i\oint_{cC_1+dC_2} A^1]}= \exp{[- 2\pi id Q.P]}=1
\end{eqnarray}
where we have used the on-shell values of the fields.

 In this case, under a  electric-magnetic duality transformation
 \begin{equation}
  \left(\begin{array}{c}
        Q\\
        P
       \end{array}\right)\rightarrow \left(\begin{array}{cc}
                                      1 & b\\
 				     0 & 1
                                     \end{array}\right)\left(\begin{array}{c}
        Q\\
        P
       \end{array}\right)
 \end{equation}
 we can bring the T-duality invariant $Q.P$ to $Q'.P=\nu$ keeping $P^2$ unchanged with $Q'^2 = 2n$ such  that  $\nu= 0$ when  $\Delta$ is even and $ \nu= 1$ when $\D$ is odd. Thus, the phase (\ref{WL phase}) becomes
 \begin{equation}
 2\pi i \frac{\D}{4}\frac{d}{c} \; \
 \end{equation}
 with $ \D = 4n - \nu^{2}$ for  $ n \in \mathbb{Z}$ and $\nu = 0, 1$,  in perfect agreement with the first phase 
 of (\ref{gener Kloosterman}).

\subsection{Phases from $SU(2)_{R}$ Chern-Simons  Terms}\label{CS phase}

The Killing spinor $\xi$ on $AdS_2\times S^2$ associated with the supercharge used for localization has the schematic form
\begin{equation}
 \xi\sim e^{\frac{i}{2}(\tilde{\theta}+\phi)}\eta
\end{equation}where $\phi$ is the azimuthal coordinate on the $S^2$ and $\eta$ is a Dirac spinor independent of both  $\tilde{\theta},\phi$.   Under the  orbifold identification (\ref{orbifold ident}) this spinor  picks a non-trivial phase and would not be invariant \cite{Banerjee:2009af}.
To preserve the supercharge we turn on an $SU(2)$ flat connection on $S^2$ so that its holonomy  along the cycle $C_1$ parametrized by $\tilde{\theta}$ gives a  compensating phase.  This together with \eqref{ARWilson} implies 
\be \label{holC1}
\oint_{C_{1}}A_R = -\frac{2 \pi i}{c}\frac{\sigma^3}{2} \, , \qquad 
\oint_{C_{2}} A_R=0 \, .
\ee
where   $A_R=\frac{i}{2} A^a\sigma^a$ is in the fundamental of $SU(2)$.

The boundary contribution to the  action (\ref{CS bnd term micro}) vanishes because   Wilson line along $C_{2}$ is fixed to zero. The bulk Chern-Simons action for the $SU(2)$ gauge field is then of the form
\begin{equation}\label{action w CS2}
 S [A_{R}] = -\frac{i k_R}{4\pi}I[{A}_R]
\end{equation}
with $k_{R} = p^{1} P^{2}/2 $ which can be obtained from the Kaluza-Klein reduction of  five-dimensional action of M-theory  \cite{Hansen:2006wu}. We thus need to compute the Chern-Simons integral \eqref{CS}
for  flat connections  subject to the boundary conditions \eqref{holC1}

This requires some mathematical machinery, which fortunately has been developed earlier  in \cite{kirk1990}  for computing  Chern-Simons invariants of Lens spaces. We review this computation, especially theorems 4.2 and 5.1 in \cite{kirk1990}. 
Our manifold of interest $\mathcal{M}(c, d)$  is topologically a solid torus $D^2\times S^1$.
We parametrize the solid torus using the coordinates $(r,x_{c},x_{n})$ and orientation $(\partial_r,\partial_c,\partial_n)$. Here $x_{c}, x_{n}\in [0,2\pi]$ are the meridian and longitude directions, that is, they parametrize the contractible and non-contractible cycles respectively, while $r\in[0,1]$ parametrizes a radial direction. A flat connection is always pure gauge
\begin{equation}\label{flat connection}
 A=-dg g^{-1},\;g\in SU(2).
\end{equation}In \cite{kirk1990} it is shown that we can always bring the gauge transformation $g$ to a form 
\begin{equation}\label{pure gauge}
 g=f(x_{c},r)e^{-\frac{i}{2}\beta \sigma^3 x_{n} }
\end{equation}where $f(x_{c},r):D^2\rightarrow SU(2)$ is a smooth function so that $f(x_{c},r)=e^{-\frac{i}{2}\alpha \sigma^3 x_{c}}$ for $r\in [1-\epsilon,1]$ and is constant in the neighborhood of the origin. The parametrization (\ref{pure gauge}) is said to be in normal form since the holonomies become diagonal in the basis chosen. One computes
\begin{equation}\label{smooth gauge connection}
 A=-\partial_c f\,f^{-1}dx_{c}-\partial_r f\,f^{-1}dr-\frac{i}{2}\beta f\sigma^3 f^{-1} dx_{n}.
\end{equation}Using the flatness condition $dA+A\wedge A=0$ one then obtains
\begin{equation}
 I[A_{R}] =-\frac{i}{2}\beta \,  \int \text{Tr}\Big(\left[\partial_c f\,f^{-1},\partial_rf\, f^{-1}\right](f\sigma^3f^{-1})\Big)dx_{c}\wedge dr\wedge dx_{n}
\end{equation}Let $\omega=\text{Tr}\left(df\wedge \sigma^3 f^{-1}\right)$. Integrating over $x_{n}$ yields precisely $d\omega$ and by Stokes' theorem we get the desired result
\begin{eqnarray}\label{CS action theorem}
I[A_{R}] = \pi i \beta\int_{C_{c}}\omega 
= -\pi i \beta\int_{0}^{2\pi}\text{Tr}(\frac{i}{2}\alpha\sigma^3\sigma^3)dx=2 \pi ^{2}\alpha\beta \, .
\end{eqnarray}
which enables us to evaluate the bulk actions given  $\a$ and $\b$. 

 Given (\ref{deg vs ind}) and \eqref{holC1},  it  is easy to compute  $\a$ and $\b$ for the $SU(2)_{R}$ gauge field:
\begin{equation}\label{WL A_R S_n}
\oint_{C_{c}}A_R = c\oint_{C_{1}}A_R+d\oint_{C_{2}}A_R = -2 \pi i\frac{\sigma^3}{2} 
\end{equation}
and
\beq
\oint_{C_{n}}A_R = a \oint_{C_{1}}A_R+ b\oint_{C_{2}}A_R=-2 \pi i\frac{a}{c}\frac{\sigma^3}{2} \, .
\eeq  
We thus conclude $\alpha=-1$ and $\beta=-a/c$. Note that $\alpha$ must be an integer because the holonomy $ \exp{(\oint_{C_c} A)}$ around the contractible cycle  must be trivial for $A$ in the vector representation. On the other hand,   $\b$ can be real.
The total  contribution from the $SU(2)_R$ action is
\begin{equation} \label{CSPhaseac}
S=  -\frac{ik_{R}}{4\pi} I[A_{R}] = -\frac{2\pi i k_{R}}{4}\frac{a}{c} \, .
\end{equation} 

\subsection{Phases from Gravitational Chern-Simons Terms}\label{Sec grav CS}

 We have seen that the $SU(2)_R$ Chern-Simons terms contribute  a phase
 \begin{equation}
\exp{\left(-2\pi i\frac{k_{R}}{4}\frac{a}{c}\right)}
 \end{equation}with $k_{R}$ the $SU(2)_R$ Chern-Simons level.  On the other hand, the microscopic counting formulae for half-BPS and quarter-BPS states in $\mathcal{N}=4$ string theory \cite{Dabholkar:2010rm}, imply that the Kloosterman sums should carry a phase proportional to the \textit{left-moving} central charge $c_L$ in the sector where supersymmetry is broken:
 \begin{equation}
\exp{\left(-2\pi i\frac{c_L}{24}\frac{a}{c}\right)}\, .
 \end{equation}
 In the boundary theory,  we have an affine $SU(2)_{R}$ current algebra with level $k_{R}$ which is related  by supersymmetry to the right-moving central charge as $c_R=6k_R$.
 However, in general,  $c_R \neq c_{L}$ which leads to a puzzle of explaining this discrepancy between the boundary and the bulk. 
 
 For the $\mathcal{N}=8$ theory, $c_L=c_R$ and  one does not encounter this puzzle. However, $\mathcal{N}=4$ actions  contain gravitational Chern-Simons terms in three dimensions \cite{Maldacena:1997de}  that are responsible for the difference between $c_L$ and  $c_R$.  We now show that these terms contribute with additional phases that precisely resolve this discrepancy between the bulk and the boundary. 
 For computing the phases using   (\ref{CS action theorem}) we need  the holonomies along the contractible and non-contractible cycles.  Their computation is considerably simpler  \cite{Gukov:2003na} if we use the Chern-Simons formulation of the three dimensional gravity which we now describe\footnote{The usual metric formulation requires  a suitable choice of coordinates which complicates the computation.}.

It is well known that the Einstein-Hilbert action\footnote{We weight the Euclidean functional integral by $e^{S}$.} with negative cosmological constant plus gauge and gravitational Chern-Simons terms can be written only in terms of the  $SL(2,\mathbb{R})_L\times SL(2,\mathbb{R})_R$ connections $\tilde A_{L,R}$ and $SU(2)_R$ connections ${A}_R$ as
\begin{equation}\label{action w CS3}
 S=-\frac{i \tilde k_L}{4\pi}I[\tilde A_L]+\frac{i \tilde k_R}{4\pi}I[\tilde A_R]-\frac{i k_R}{4\pi}I[{A}_R]
\end{equation}
with $\tilde k_L=c_L/6$ and $\tilde k_R=c_R/6$.  
As discussed in \S\ref{sec CS bnd terms} we need to add appropriate boundary terms.   
The analysis is similar to the $SU(2)$ case except that for $SL(2,\mathbb{R})$ the Wilson lines can be complex instead of imaginary. At the on-shell level both $A_{L,R}$ are flat and therefore we need to compute the Chern-Simons integral of an $SL(2,\mathbb{R})$ flat connection.

We view  $AdS_3$ as the group manifold $SO(2,2)/SO(2,1)$ so that  thermal $AdS_3$ corresponds to the identification
\beq
\mathbf{X}\sim g_L \mathbf X g_R
\eeq where $\mathbf{X}$ is an element of $SO(2,2)$ and $g_{L,R}=\textrm{diag}(e^{\pi \tau_{L,R}},e^{-\pi \tau_{L,R}})$ are elements of $SL(2,\mathbb{R})_{L,R}$. For the $SL(2,\mathbb{Z})$ family of solutions we need to complexify $\tau_{L,R}$. The left quotient is generated by the elements 
\beq\label{SL2 identification}
g_L=\left\{\left(\begin{array}{cc}
     e^{i\pi \frac{a\tau+b}{c\tau+d}} & 0\\
      0 & e^{-i\pi \frac{a\tau+b}{c\tau+d}}
    \end{array}\right),\; \left(\begin{array}{cc}
     e^{i\pi} & 0\\
      0 & e^{-i\pi}
    \end{array}\right)\right\}
\eeq 
 Similarly we could have considered Euclidean $AdS_3$ under the indentification $\mathbf{X}\sim\rho(\tau) \mathbf{X} \rho^{\dagger}(\tau)$ with $\rho(\tau)$ a diagonal subgroup of $SL(2,\mathbb{C})$ and take $\tau$ and $\bar{\tau}$ independent.  
The first element in (\ref{SL2 identification}) corresponds precisely to the holonomy of $\tilde A_{L}$ around the non-contractible cycle whereas  the second element corresponds to the holonomy of $\tilde A_{L}$ around the contractible cycle \cite{Castro:2011xb}. In the Euclidean case the right  quotient is  generated by elements that are just hermitiean conjugates $g_{R} = g_{L}^{\dagger}$ and correspond similarly to the holonomies of $\tilde A_{R}$ around the non-contractible and contractible cycles.   We can  read off easily the corresponding Wilson lines in the extremal limit $\bar{\tau}\rightarrow\infty$  
  \bea
\oint_{C_{n}}\tilde A_L =  2\pi i  \frac{a\tau+b}{c\tau+d} \frac{\sigma^3}{2} & \, ,& \qquad \oint_{C_{n}}\tilde A_R =  - 2\pi i  \frac{a}{c} \frac{\sigma^3}{2}\, ; \\
\oint_{C_{c}}\tilde A_L =  2\pi i   \frac{\sigma^3}{2} & \, , & \qquad \oint_{C_{c}}\tilde A_R =  - 2\pi i   \frac{\sigma^3}{2}\, .
\eea
Given these Wilson lines it is easy to determine the Wilson lines along the cycles $(C_{1},C_2)$:
\bea
\oint_{C_{2}} \tilde A_L= 2\pi i \, \frac{1}{c\tau+d} \frac{\sigma^3}{2} &\, ,& \qquad \oint_{C_{2}} \tilde A_R=0 \,  \label{AL} \, ; \\
\oint_{C_1}\tilde A_L=2\pi i \, \frac{\tau}{c\tau+d} \frac{\sigma^3}{2}&\, ,& \qquad \oint_{C_1}\tilde A_R=- 2\pi i \frac{1}{c} \frac{\sigma^3}{2} \, .
\eea
As explained before, the boundary conditions fix the Wilson lines along the $C_{2}$ cycle.  
Note that first condition in \eqref{AL} is nothing but the requirement that the radius $R$ in the metric (\ref{metric ads2xs1}) is held fixed. As discussed in \S\ref{Sec Wl + CS phase}  the Wilson lines along the $C_{1}$ cycle in  the classical problem are determined by  regularity. 
Given the Wilson lines we can use the result (\ref{CS action theorem}) to compute
\beq
\frac{1}{4\pi}I[\tilde A_L]=\frac{\pi}{2}\frac{a\tau+b}{c\tau+d}\, , \qquad \frac{1}{4\pi}I[\tilde A_R]=\frac{\pi}{2}\frac{a}{c} \, ;
\eeq 
together with the boundary terms
\beq
\frac{1}{4\pi} \int \text{Tr}\tilde A_{L1}\tilde A_{L2}=-\frac{\pi}{2}\frac{\tau}{(c\tau+d)^2} \, , \qquad \frac{1}{4\pi} \int \text{Tr}\tilde A_{R1}\tilde A_{R2}=0 \, .
\eeq

Putting it together,  we compute the total action  (\ref{action w CS3}) including the  boundary terms. The $SL(2)_R$ and $SU(2)_R$ terms cancel because by supersymmetry  $\tilde{k}_R=k_R$ and we  obtain\begin{equation}\label{CS on-shell}
 S=-2\pi i\frac{c_L}{24}\frac{a\tau+b}{c\tau+d}\, + \,  2\pi i\frac{c_L}{24} \frac{\tau}{(c\tau+d)^2} \, .
\end{equation}
where we have used $c_{L} = 6 \tilde k_{L}$. 

Using the equalities
\begin{equation}\label{equality 1}
 \frac{a\tau+b}{c\tau+d}=\frac{a}{c}-\frac{1}{c(c\tau+d)}\, \qquad \frac{\tau}{(c\tau+d)^2}=\frac{1}{c(c\tau+d)}-\frac{d}{c}\frac{1}{(c\tau+d)^2}
\end{equation}
we write (\ref{CS on-shell})  as
\begin{eqnarray}\label{CS action extremal lim}
 S&=&-2\pi i\frac{c_L}{24}\frac{a}{c}\, + \, 2\pi i\frac{c_L}{12}\frac{1}{c(c\tau+d)}\, - \,2\pi i\frac{c_L}{24}\frac{d}{c}\frac{1}{(c\tau+d)^2} \, .
\end{eqnarray} 
Using \eqref{R} we can reexpress it in terms of the radius $R$
\be\label{SR}
S&=& - \, \left ( 2\pi \frac{c_L}{12}\frac{R}{c} \right) \, + \, \left( -2\pi i\frac{c_L}{24}\frac{a}{c}\,   + \,2\pi i\frac{c_L}{24}\frac{d}{c}R^{2} \right) \, .
\ee
The term in the first bracket is real and can be identified with classical entropy of the black hole reduced by a factor of $c$ if
\be
R = - {2}/{\phi^{0}_{*}} \, .
\ee
This gives the correct entropy $S_{BH}$ for these black holes from the real term in \eqref{SR}.
We can therefore rewrite (\ref{CS action extremal lim}) as
\begin{equation}\label{Ren action frm CS}
 S= \frac{1}{c}S_{BH}  + 2\pi i (\text{phase} ) \, .
\end{equation}
where $S_{BH}$ is the  on-shell classical black hole entropy. 
Our analysis in this section is for the \textit{on-shell} theory. 
We believe that this result can be extended to the off-shell theory in which case we would replace $S_{BH}$  by the off-shell renormalized action (\ref{Sren}) whose extremum value gives $S_{BH}$ . Since the phase of our interest is topological coming from terms supported only at the boundary we expect that it will be the same also in the  off-shell extension of our results. 

We see now that the coefficient multiplying $a/c$ in \eqref{SR} is in fact proportional to $c_L/24$ which resolves the puzzle posed in  the beginning.  For half-BPS states in~$\CN=4$ theory 
we have\footnote{In our conventions the radius $R$ is negative.}
\be
c_{L }= 24 p^{1}\, , \qquad R^{2}  = -\frac{q^{0}}{p^{1}} \, .
\ee 
Hence,  the total phase from this sector by exponentiating the imaginary part in \eqref{SR} is
\be
e^{2\pi i \left(-q^{0}  \frac{d}{c} -  p^{1} \frac{a}{c}  \right)}
\ee
For the one-eighth BPS states in the~$\CN=8$ theory, we have
\be\label{on-shell val R}
c_{L }=  6 k_{L} \, , \qquad  R^{2 } = \frac{\D}{k_{L}^{2}} \, ,
\ee
where we have used  $p^{1}=1$  so that\footnote{This ensures $(4,4)$ supersymmetry for the near horizon geometry  and implies $k_{L} = k_{R}$.}
\be
k_{L} =  P^{2}/2 \, .
\ee
Hence, the total phase from this sector by exponentiating the imaginary part in \eqref{SR} is
\be\label{covphase}
e^{2\pi i \frac{d}{c}\frac{\D}{4 k_{L}}}  \; 
e^{-2\pi i \frac{a}{c} \frac{k_{L}}{4}}.
\ee
Recalling that $k_{L}=k_{R}$ in the~$\CN=8$ theory, we see that the phase~\eqref{covphase} reproduces
the product of the phases~(\ref{CSPhaseac}) and~(\ref{WL phase}) derived previously.

\section{The Multiplier System and Duality Invariance \label{Multiplier}}

The  phases that we have worked out thus far   are common to all known models with at least eight supercharges.  We now turn to the computation of the multiplier matrix $M(\gamma)_{\mu\nu}$ which is crucial for invariance of the Rademacher sum under the elements $\Gamma_{\infty}\in SL(2,\mathbb{Z})$. This ensures that we are not overcounting geometries.

To set up the computation  we first describe an explicit representation of the multiplier matrix in section \ref{multiplier sect} following the work of \cite{Lisa92}. In \S\ref{Physical} we give a physical interpretation of this representation in terms of  $SU(2)_L$ Chern-Simons path integral. Recall that the degeneracy of one-eighth BPS black holes in 4d  is the same as the  degeneracy of one-quarter BPS black holes in 5d.  By the 4d-5d lift the charge $q_{1}$ in the presence of a KK-monopole with charge $p^{1}=1$ can be identified with the spin of this 5d black hole.   This black hole is an excited state of the D1D5P Strominger-Vafa black string in six dimensions with near horizon $AdS_{3} \times S^{3}$. As a result,  the near horizon theory has $(4, 4)$ supersymmetry and $SU(2)_{R} \times SU(2)_{L}$ R-symmetry to start with. This is a geometric symmetry of the $S^{3}$ which is broken to $SU(2)_{R} \times U(1)_{L}$ by the spin of the black hole. 

 While localization fixes the holonomy of the $SU(2)_R$ flat connection on both contractible and non-contractible cycles, for the $SU(2)_L$ sector there is no such restriction and we have to sum over the holonomies. The Chern-Simons action with boundary terms for these flat connections combines  to reproduce the multiplier matrix $M(\gamma)$ up to a non-trivial phase which can be related to a choice of framing  as we discuss in \S\ref{Knot}.

\subsection{An Explicit Expression for the Multiplier System}\label{multiplier sect}

The matrix $M(\gamma)$ realizes a representation of $SL(2,\mathbb{Z})$ in the space of  vector-valued modular forms. It appears in the generalized Kloosterman sum via its inverse.  It is easy to determine the matrix $M^{-1}(\gamma)$ for the $S,T$ generators of $SL(2,\mathbb{Z})$ from the definition \eqref{thetadef} and the transformation properties of the theta functions:
\begin{equation}\label{images ST}
 \tilde{T}\equiv M^{-1}(T)=\left(\begin{array}{cc}
                  1 & 0\\
		  0 & i
                 \end{array}
\right),\;\;\tilde{S}\equiv M^{-1}(S)=\frac{e^{-\frac{\pi i}{4}}}{\sqrt{2}}\left(\begin{array}{cc}
                  1 & 1\\
		  1 & -1
                 \end{array}
\right)
\end{equation}
To find $M^{-1}$ for a general  element $\gamma$, we can use its \emph{continued fraction expansion} $[m_1,\ldots,m_t]$ 
\begin{equation}
 \gamma=T^{m_t}S\ldots T^{m_1}S 
\end{equation}and iteratively compute the matrix $M^{-1}$ via the images (\ref{images ST})
\begin{equation}\label{ctnd fract expansion}
 M^{-1}(T^{m_t}S\ldots T^{m_1}S)_{\mu_t\mu_1}=(\tilde{T}_{\mu_t\mu_t})^{m_t}\tilde{S}_{\mu_t\mu_{t-1}}(\tilde{T}_{\mu_{t-1}\mu_{t-1}})^{m_{t-1}}\tilde{S}_{\mu_{t-1}\mu_{t-2}}\ldots\tilde{S}_{\mu_{2}\mu_{1}}
\end{equation}
Even though this is conceptually straightforward, it is  nontrivial to obtain a useful explicit expression from this formula. 
 Fortunately, this problem  has already been addressed  by Jeffrey \cite{Lisa92} in the context of knot theory to construct the Chern-Simons partition function using the surgery formula of Witten \cite{Witten:1988hf}.  In the following we use her results to construct an explicit representation for the multiplier matrix which is particularly  amenable to a simple physical interpretation.  This connection to knot invariants via Chern-Simons theory  is not accidental as we explain in \S\ref{Knot}.

Jeffrey uses the representations of $S,T\in SL(2,\mathbb{Z})$  given by
\begin{equation}\label{rep Lisa92}
 S_{jl}=\sqrt{\frac{2}{r}}\sin\left(\frac{jl\pi}{r}\right);\;\; T_{jl}=\zeta^{-1}e^{2\pi i\frac{j^2}{4r}}\delta_{jl},\;\zeta=e^{\frac{\pi i}{4}}
\end{equation}with the basis $j,l$ running from $1$ to $r-1$, and $r$ is the ``renormalized'' Chern-Simons level, that is, $r=k+2$ for $SU(2)$ gauge group. Since we have $k=1$ then $r=3$, and therefore the representation is two dimensional with $j,l=1,2$. Identifying the basis $j,l$ with our basis $\nu, \mu$ in (\ref{images ST}) as $j=\nu+1$ we observe that $S_{jl}$ and $T_{jl}$ in (\ref{rep Lisa92}) are the same as $M^{-1}(S,T)$ in (\ref{images ST}) up to an overall phase, that is,
\begin{equation}
 \tilde{S}=e^{-\frac{\pi i}{4}}S_{jl},\;\;\tilde{T}=e^{\frac{\pi i}{12}} T_{jl}.
\end{equation}
Our general representation for (\ref{ctnd fract expansion})  has an additional phase relative to the result in  \cite{Lisa92}:
\begin{equation}\label{phase vs lisa}
 M^{-1}(\gamma)_{\nu\mu}=e^{\frac{\pi i}{12}\left(\sum_{l=1}^t m_{l}-3t\right)}\tilde{U}_{\mu+1, \nu+1}
\end{equation}where $\tilde{U}$ is given by Proposition 2.7 in \cite{Lisa92}
\begin{equation}\label{tildeU}
 \tilde{U}(\gamma)_{j l}=i\frac{\text{sign}(c)}{\sqrt{2r |c|}}e^{-\frac{i\pi}{4}\Phi(\gamma)}\sum_{\epsilon=\pm}\sum_{n=0}^{c-1}\epsilon \,e^{\frac{i\pi}{2rc}[dj^2-2j(2rn+\epsilon l)+a(2rn+\epsilon l)^2]},\, \end{equation}with $\Phi(\gamma)$ the Rademacher phi function defined as
\begin{equation}\label{RadPhi}
 \Phi\left[\begin{array}{cc}
                  a & b\\
		  c & d
                 \end{array}
\right]=\frac{a+d}{c}-12\text{sign}(c)s(a,|c|),
\end{equation}and $s(a,c)=s(d,c)$ is the Dedekind sum:
\begin{equation}
 s(a,c)=\frac{1}{4c}\sum_{j=1}^{c-1}\cot\left(\frac{\pi j}{c}\right)\cot\left(\frac{\pi ja}{c}\right),\;c>0.
\end{equation}As shown in \cite{Lisa92} (Lemma 3.1), for an $SL(2,\mathbb{Z})$ matrix with $-c/a>1$, which is in agreement with the choices for $(c,d,a)$ in the Kloosterman sum,  the continued fraction expansion has a particular form
\begin{equation}
 \gamma=ST^{m_{t-1}}S\ldots T^{m_1}S,\;m_i\geq 2,
\end{equation}with the value of the Rademacher phi function given in this case by
\begin{equation}
 \Phi(\gamma)=\sum_{l=1}^{t}m_l-3(t-1).
\end{equation}Given these results, the relative phase in (\ref{phase vs lisa}) is computed to give
\begin{equation}
 e^{\frac{\pi i}{12}\left(\sum_{l=1}^t m_{l}-3t\right)}=e^{\frac{i\pi}{12}(\Phi(\gamma)-3)}.
\end{equation}
This phase is reminiscent of the framing correction in knot theory. We discuss this connection in 
\S\ref{Knot}.  The final result for the multiplier matrix is then
\begin{equation}\label{multiplier matrix}
  M^{-1}(\gamma)_{\nu\mu} = C \sum_{\epsilon=\pm}\sum_{n=0}^{c-1}\epsilon \, e^{\frac{i\pi}{2rc}\left[d (\n +1)^2-2(\n +1)(2rn+\epsilon (\m +1))+a(2rn+\epsilon (\m +1))^2\right]}
\end{equation}
where the prefactor is
\be
C := i\frac{\text{sign}(c)e^{-\frac{\pi i}{4}}}{\sqrt{2r |c|}}e^{-\frac{i\pi}{6}\Phi(\gamma)} \, .
\ee
It can be shown from this expression that the generalized Kloosterman sum  depends only on the equivalence class $\Gamma_{\infty}\backslash \gamma/\Gamma_{\infty}$ where  we need the matrix elements $M^{-1}(\gamma)_{\nu 1}$ for $\nu=0,1$.

\subsection{A Physical Interpretation from $SU(2)_{L}$ Chern-Simons Theory }\label{Physical}

To compute the boundary contribution \eqref{CS bnd term micro} for the $SU(2)_{L}$ gauge field, we need the Wilson lines along the cycles $C_{1}$ and $C_{2}$. These in turn determine the Wilson lines along the cycles $C_{n}$ and $C_{c}$ parametrized by  $\alpha$ and $\beta$ in \eqref{ab} which can be used to compute the bulk action \eqref{CS} using the theorem \eqref{CS action theorem}. 

As explained in \S\ref{sec CS bnd terms}, the Wilson line $\delta$ around the cycle $C_{2}$ is fixed by the boundary condition. Because $Q.P$ is the spin $J_{L}$ of the black hole in 5d under $SU(2)_{L}$, this implies
\be\label{defdelta}
\delta:= \oint_{C_{2}} A  = 2\pi i \frac{\sigma^3}{2}J_{L }/k_{L}= \pi i\sigma^3\frac{Q.P}{k_{L}}
\ee
The Wilson line $\gamma$ around the cycle $C_{1}$ is  determined by demanding that the holonomy around the  around the contractible cycle in the vector representation must be trivial. 
Since we are working in the fundamental representation and since $C_{c} = c C_{1} + d C_{2}$ we require
\begin{equation}\label{cond SU(2)_L} 
 \exp{\left(c \oint_{C_{1}} A + d \oint_{C_{2}} A  \right)} = \pm \mathbb{I}
  \end{equation}
Therefore,  using \eqref{defdelta} and defining  $\nu$ by
\be
Q.P=\nu+2lk_{L}, \,\, l\in\mathbb{Z}, 
\ee 
we  get
\beq
\gamma := \oint_{C_{1}} A  = \frac{1}{c}\left(\lambda+2n-d\frac{\nu}{k_{L}}\right),\qquad n\in \mathbb{Z}/c\mathbb{Z} \, , \quad \lambda = 0,\pm 1 \, .
\eeq 
To determine the conditions $n\in \mathbb{Z}/c\mathbb{Z}$ and $\lambda = 0,\pm 1$, we need to examine the symmetries of the exponential of the Chern-Simons action. As a matter of fact it would be enough to consider $\lambda=0,1$ so that (\ref{cond SU(2)_L}) is satisfied. However, as we see next, the exponential of the Chern-Simons action is different for $\lambda=\pm 1$ and therefore we should consider this other possibility. The Wilson lines are computed to give
\begin{eqnarray}
\alpha &:=& \oint_{C_c} A=\pi i\sigma^3\left(c\gamma+d\frac{Q.P}{k_{L}}\right)=\pi i\sigma^3\left[\lambda+2n+2dl\right]\\
\beta &:=& \oint_{C_n} A=\pi i\sigma^3\left(a\gamma+b\frac{Q.P}{k_{L}}\right)=\pi i\sigma^3\left[\frac{a}{c}(\lambda+2n)-\frac{\nu}{ck_{L}}+2lb\right]
\end{eqnarray}

Given $\gamma$ and $\delta$, the boundary action (\ref{CS bnd term micro}) becomes
\beq
\frac{ik}{4\pi}I_{b}(A_{L}) =\frac{i\pi}{2k_{L}}\frac{d}{c}\nu (\nu+2lk_{L})-\frac{i\pi}{2c}(2n+\lambda)(\nu+2lk_{L}) \, ,
\eeq 
and given  $\a$ and $\b$, the bulk Chern-Simons action \eqref{CS} becomes
\beq
\frac{ik}{4\pi}I[A_L]=k_{L}\frac{i\pi }{2}\left[\frac{a}{c}(2n+\lambda)^2-\frac{\nu}{ck_{L}}(2n+\lambda)+\frac{2l}{c}(2n+\lambda)-\frac{d}{c}\frac{\nu}{k_{L}}2l+\text{mod}(4)\right]
\eeq
Putting both terms together we arrive at 
\begin{equation}
 S_{CS}=\frac{i\pi}{2k_{L}}\frac{d}{c}\nu^2-i\pi\frac{\nu}{c}\left(2n+\e\frac{\mu}{k_{L}}\right)+k_{L}\frac{i\pi}{2}\frac{a}{c}\left(2n+\e\frac{\mu}{k_{L}}\right)^2+\text{mod}(2\pi i)
\end{equation}  
where we define $\l = \e \mu /k_{L} $ and $\e = \pm 1$. 
When we exponentiate this expression and sum over sectors  we get precisely the Kloosterman phase under the sums in (\ref{multiplier matrix}) except that there the constants $k$ and $\nu,\mu$ are renormalized to $k+2, \nu+1, \mu+1$ respectively.  

It is well known that the Chern-Simons level $k_{L}$ gets renormalized. In  our computation $\nu, \mu$ could possibly get renormalized if the boundary  conditions on the holonomies get shifted appropriately. 
The prefactor $C$   of the multiplier matrix (\ref{multiplier matrix}) does not follow from our classical evaluation
but it could possibly arise from one-loop corrections of the localization action. Indeed, in the context of localization of Chern-Simons theory on Lens spaces, such a phase  proportional to the Rademacher phi function is known to arise precisely in this manner  \cite{Beasley:2009mb,Kallen:2011ny}. 
There is a natural connection of our computation  to  knot theory on Lens space   and Witten's surgery formula which we elaborate in \S\ref{Knot}. An extension of the results \eqref{CS on M(c,d)} and \eqref{Witten CS} from knot theory could explain both the renormalization of $k_{L}, \n, \m$ and the prefactor $C$. 

There is one puzzle that we do not fully understand. The $U(1)$ part of the $SU(2)_{L}$ gauge field  can be identified with the $U(1)$ abelian gauge fields that couples to the charge $q_{1}$. This can be seen using for example by using 4d-5d lift in which electric charge $q_{1}$ in 4d in the presence of a $KK$ monopole of charge $p^{1}=1$ gets identified with the spin $J_{L}$.
For charge configurations with $q_{1} \neq 0$ there seems to be a puzzle because this  $U(1)$ gauge field is apparently counted twice in computing  both the  phase in \S\ref{Sec Wl + CS phase} as well as the multiplier system in this subsection. However, note that the action used in this subsection is obtained by truncating a 6d theory on an $S^{3}$, whereas the action used in \S\ref{Sec Wl + CS phase} is obtained by first reducing on a circle and then truncating on  an $S^{2}$. It would be necessary to understand the precise map between these truncations to resolve this puzzle.

\subsection{Duality Invariance and Kloosterman Sums \label{Duality}}

We now analyze the total phase after summing over sectors assembling all terms. Duality invariance of the resulting answer is not immediately obvious and requires nontrivial identities for the Kloosterman sum.

For the half-BPS black holes\footnote{These black holes are `small'  \cite{Dabholkar:2005dt} in that  their horizon area is of order one in string units. As a result, one does not really have a geometric picture. However,  the localization solutions depend essentially on the symmetries of $AdS_{2}$ and the Kloosterman sum is topological which can partially justify our supergravity reasoning which in any case seems to lead to a striking agreement. Similar phases are expected to be relevant  for large black holes.} there is no multiplier system and duality invariance is relatively easy to prove. The total phase 
 is
\be
\sum_{\substack{ d\in \mathbb{Z}/c\mathbb{Z}\\
                  d a =1\,\text{mod}(c)}} e^{2\pi i \left(-q^{0}  \frac{d}{c} -  p^{1} \frac{a}{c}  \right)}
\ee
For $p^{1}=1$ this is in perfect agreement with \eqref{kloos1} with $N= -q^{0}$.

For  general $p^{1}$, we get the classical Kloosterman sum
$
Kl(-q_{0}, -p^{1},  c) \, 
$
defined in \eqref{classical Kl}. At first sight it does not seem duality-invariant. However,  using the fact that $q_{0}$ and $p^{1}$ are relatively prime, and choosing $p= p^{1}, n= -q^{0}, m= -1$ 
in \eqref{kloosrel} we get
\be\label{iden}
Kl(-q_{0}, -p^{1},  c) = Kl(-p^{1}q_{0}, -1,c)
\ee
in perfect agreement with the duality invariant formula \eqref{kloos1}.    

For the one-eighth BPS black holes we choose $k_{L}=1$ and the total phase is
\bea
\label{kloosfinal}
\sum_{\genfrac{}{}{0pt}{}{-c \leq d< 0;}{(d,c)=1}} e^{2\pi i \frac{d}{c} (\D/4)} \; M^{-1}(\gamma_{c,d})_{\nu 1} \; 
e^{2\pi i \frac{a}{c} (-1/4)} \qquad  \,   \,    \qquad  
\eea

Considerations of duality invariance thus suggest interesting identities for the Kloosterman sum. For example, for a general $k_{L}$,  it should be possible to obtain the duality invariant answer as above. However, it would require a generalization of the identity \eqref{iden} for the generalized Kloosterman sum which to our knowledge is not known in the mathematics literature.
Another important classical identity is the Selberg identity
\be
Kl(n, m, c)=\sum_{s|gcd(n, m, c)}s \,Kl\left(n/s, m/s, c/s\right) \, .
\ee
A generalization of this identity  could play a role in the context of black holes with nonprimitive charge vectors \cite{Maldacena:1999bp,Sen:2008sp,Banerjee:2008pu,Banerjee:2008pv, Dabholkar:2008zy}. 

\subsection{Relation to Knot Theory \label{Knot}}

Several ingredients that we have encountered in our computation of  the Kloosterman sums are reminiscent of knot theory in Lens space $L(c, d)$. This is not a coincidence and follows from the fact  that Chern-Simons theory on the Dehn-filled solid torus can be related to the  surgery formula  of Witten \cite{Witten:1988hf}  for the unknot in the Lens space. 

Let us recall a few facts about  the surgery formula. Any three manifold $M'$ without boundary can be obtained by  Dehn surgery on a link $\mathcal{L}$ on another manifold $M$. This procedure consists in removing a tubular  neighborhood $\text{Tub}(\mathcal{L})$ in $M$ and glueing  it back, after an $SL(2,\mathbb{Z})$ diffeomorphism of the boundary of $\text{Tub}(\mathcal{L})$.  

As an example, consider  Lens space which can be constructed as follows. Consider $M= S^2\times S^1$ which is  the union of two solid tori $\mathbf{X}_1=D^2\times S^1$ and $\mathbf{X}_2=D^2\times S^1$ glued at the boundary by the identity element. This means that we identify the boundary cycle of the first disk with the boundary cycle of the second disk and the same for $S^1$. On the other hand, if we  glue the boundary  of the first disk with the circle $S^1$ of the second solid torus and vice-versa, then we obtain the manifold $M' = S^3$. This corresponds to a Dehn twist by the element  $S$ of $SL(2, \mathbb{Z})$.  The Lens space $L(c, d)$ is obtained  as a union of the two tori after a general  Dehn twist by an element $\gamma$.  To make contact with the surgery described earlier, we see that $M = S^2\times S^1$, $M' = L(c,d)$,  $X_{1} = \text{Tub} (\mathcal{L})$ for the unknot $\mathcal{L}$ which  is glued back to $X_{2} = M \backslash X_{1}$ after an $SL(2,\mathbb{Z})$ twist.  In this case, $X_{2}$ is also a solid torus and to be consistent with our notation in the rest of the paper, we perform the Dehn twist on the boundary of $X_{2}$. To be more precise take $C_1$ to be cycle on $\partial X_2$ that becomes non-contractible in $X_2$ and $C_2$ the cycle that becomes contractible. After the twist it is the cycle $cC_1+dC_2$ that becomes contractible and $aC_1+bC_2$ is the non-contractible cycle. Note that in this case the untwisted  $X_{2}$  is  nothing but the thermal $AdS_{3}$ and the twisted $X_{2}$ are the orbifolds $\mathcal{M}(c, d) $.

The expectation value of a Wilson line $W(\mathcal{R}_{\a}) $ in representation $\mathcal{R}_{\a}$ along the link $\mathcal{L}$ in $M'$ for   Chern-Simons theory with  group $G$ and level $k$ is defined as
\beq \label{Jones}
Z(M',\mathcal{R}_{\alpha}) :=\int_{M'}  W(\mathcal{R}_{\a}) \, e^{ \frac{ik}{4\pi}I(A) }
\eeq
When $M'=S^3$ the quantity $Z(S^3,\mathcal{R}_{\alpha})$ is proportional to the Jones polynomial \cite{Witten:1988hf}.
Henceforth, we take $G=SU(2)$ and  $M' = L(c, d)$  constructed by Dehn surgery as above. In this case the  path integral \eqref{Jones} can be related to a path integral on $\mathcal{M}(c, d)$  \cite{Rozansky:1994qe} as follows. The integration  on the $\text{Tub}(\text{unknot})$ produces a delta function in the space of boundary fields. After integrating  over these boundary fields it imposes a nontrivial boundary  condition for the remaining integral on $X_{2} =  \mathcal{M}(c,d)$:
\be\label{bdrysurgery}
\oint _{C_2}A =2\pi i\frac{\sigma^3}{2} \frac{j }{r},  \;\;  j=1\ldots r-1 .
\ee 
We conclude
\beq\label{CS on M(c,d)}
Z(L(c, d),\mathcal{R}_{j})=\int_{\mathcal{M}(c,d)} e^{ \frac{ik}{4\pi} \left[I(A)  + I_{b}(A) \right]} \, .
\eeq
where the path integral on the right is subject to the boundary condition \eqref{bdrysurgery}.
We have $k=1$ and $r=3$, so $j$  takes two values: $j=1$ for the trivial representation and $j=2$ for the fundamental.

Using surgery, Witten relates \cite{Witten:1988hf} the  path integral on  $M'$ to the path integral on  $M$:
\beq\label{Witten CS}
Z(M',\mathcal{R}_{j})=e^{i\phi_{fr}}\sum_{l=1}^{r-1}Z(M,\mathcal{R}_{l}) {V}^{(c,d)}_{l j}
\eeq
where  $V^{(c,d)}_{lj}$ is a representation of the $SL(2,\mathbb{Z})$ Dehn twist in the Chern-Simons Hilbert space.  The phase $i\phi_{fr}$ is a framing correction\footnote{A framing of a knot is the analogue of  point-splitting regularization  in physics \cite{Witten:1988hf} necessary for computing the self-linking number. This regularization is not ``topological'' and  the invariant depends on the choice of the framing. When the knot invariants are computed in the canonical framing there is a phase called framing correction relative to the invariants on manifolds obtained by successive Dehn-surgeries \cite{Witten:1988hf,Rozansky:1994qe}. }  which is given by
\be\label{frame correction}
\phi_{fr}=\frac{\pi}{12}\frac{r-c_V}{r}\text{dim}G\left[\Phi[\gamma]-3\text{sign}(\frac{d}{c}+\nu)\right]
\ee
where  $c_{V}$ is the dual Coxeter number, $\text{dim} G$ is the dimension of the group,   $\nu$ is the self-linking number of the knot $\mathcal{K}$, and $\Phi[\gamma]$ is the Rademacher phi function introduced in \eqref{RadPhi}.

It is known that $Z(S^2\times S^1,\mathcal{R}_{l})=\delta_{l,1}$ and the construction of  $L(c, d)$ by Dehn surgery described earlier corresponds to choosing $\mathcal{L}$ to be the unknot.  We can then conclude
\be\label{LtoM}
Z(L(c,d),\mathcal{R}_{j})= e^{i\phi_{fr}}{V}^{(c,d)}_{1j}
\ee
By computing explicitly this matrix element in \cite{Lisa92}, Jeffrey was able to  obtain exact result for the path integral \eqref{Jones} for Lens space. Moreover,  (\ref{CS on M(c,d)}) implies that the Chern-Simons integral on $\mathcal{M}(c,d)$  with the boundary conditions (\ref{bdrysurgery}) must equal $e^{i\phi_{fr}}V^{(c,d)}_{1j}$.

This makes contact with  the black hole problem if the matrix  $V^{(c,d)}$ can be identified with $\tilde U^{(c,d)}$. We see that up to the $\text{sign}(\frac{d}{c}+\nu)$ in (\ref{frame correction}), our multiplier system is precisely the matrix in (\ref{Witten CS}). It  seems though that in the black hole context more general boundary conditions  are relevant because we need the matrix elements $\tilde U_{21}^{(c, d)}$ and $\tilde U_{22}^{(c, d)}$. We have seen in \S\ref{Sec grav CS} that the phases in the Kloosterman sum can be computed from the \textit{on-shell} $SL(2)$ Chern-Simons theory whereas the multiplier matrix comes from the $SU(2)_{L}$ Chern-Simons theory. It raises the intriguing possibility that the \textit{entire} Kloosterman sum,  which has deep connections in number theory, can perhaps be related to a topological computation in knot theory for $SL(2) \times SU(2)$ and appropriate Wilson lines, even in the general case \eqref{gener Kloosterman}.

 \section{An Assessment \label{Assess}}
 
The program of  computing the fully quantum corrected entropy of supersymmetric black holes has evolved considerably over the past few years. A  number of difficult hurdles have been overcome and there has been some important progress that brings us very close to this goal. We now give an assessment  of the status.

 \subsection{Solved Problems}
 
   $\bullet$  \textit{Choice of the ensemble}:
Even though the Bekenstein-Hawking entropy is independent of the choice of the ensemble, the finite size subleading corrections to the entropy depend sensitively on the choice of the ensemble.  The  $AdS_{2}$ boundary conditions imply that the \textit{microcanonical} ensemble is a natural choice \cite{Sen:2008yk,Sen:2008vm}. As we have seen, the boundary conditions set by this ensemble play an important role in   our computations of the Kloosterman sums.

\vspace{2mm} 
\noindent $\bullet$  \textit{Relation between  index and degeneracy}:
  Black hole entropy must correspond to the total degeneracy by  Boltzmann relation.
 On the other hand, the microscopic degeneracies are usually computed using a spacetime index.  It is not clear \textit{a priori} why the two should agree. There has been considerable confusion on this issue.  The $AdS_{2}$ boundary conditions explain that  the index equals the degeneracy near the horizon of a single black hole \cite{Sen:2009vz}. In general there are additional  contributions to the  spacetime index from degrees of freedom outside a single  horizon  \cite{Banerjee:2009uk,Jatkar:2009yd} which can complicate the comparison \cite{Denef:2007vg, Dabholkar:2010rm}. However, for the two examples considered in this paper such `hair' degrees of freedom are absent which allows for a direct comparison.
 
\vspace{2mm}
\noindent $\bullet$  \textit{Localization in supergravity}:
The formal supergravity functional integral could be evaluated  only because localization   reduces it to a finite-dimensional integral. The localizing instantons have simple analytic expressions  for   supergravity coupled to vector multiplets \cite{Dabholkar:2010uh}. They are \textit{universal} because they depend only on the off-shell supersymmetry transformations that are independent of  compactification and the  physical action.  It has been shown subsequently that  other modes of 
the supergravity multiplet do not play a role in localization~\cite{Gupta:2012cy}  and hence  these are the most general  solutions. To our knowledge this is the first application of localization  in a gravitational context. 
It may appear that the metric does not play any role in localization however that is because in the superconformal formalism, we have chosen a gauge  that trades off the conformal factor of the metric for a compensating scalar 
field. Thus,  the conformal mode of the \textit{physical} metric does have a nontrivial profile for the localizing solution. 

\vspace{2mm} 
\noindent $\bullet$  \textit{Contribution from D-type terms}:
In the physical action we only considered  chiral superspace integrals governed by a prepotential (F-type terms) and ignored  full superspace integrals 
(D-type terms). 
It was shown in \cite{deWit:2010za} using the enhanced supersymmetry of the near horizon geometry that the on-shell classical 
entropy of BPS black holes in $\CN=2$ theories does not receive any contributions from a large class of D-terms. 
However, in the context of localization, one must show that  D-terms do not contribute to the \textit{off-shell} renormalized action evaluated on the 
universal localizing solutions. 
This  was shown in~\cite{Murthy:2013xpa} for the class of D-terms considered in \cite{deWit:2010za}, 
thus providing a justification for  our analysis that ignores the D-type terms.

 \vspace{2mm}
\noindent $\bullet$  \textit{Origin of nonperturbative  corrections and phases}:
The results of this paper make clear that the supergravity functional integral is fully capable of reproducing not only the perturbative corrections to the Bekenstein-Hawking entropy but also  the nonperturbative corrections including all  intricate details of the Kloosterman Sum. In the broader context of quantum holography,  the finite charge corrections considered here correspond to finite $N$ corrections that  are perturbative (expansion in $1/N^{2}$) and nonperturbative (expansion in  ${e^{-N^{2}}}$). It is remarkable that the quantum gravity functional integral in this case can capture all these corrections.
 
 \subsection{Open Problems}

  \vspace{2mm}
\noindent $\bullet$  \textit{Contribution from hypermultiplets and gravitini}:
We have analyzed the $\CN=8, 4$ theories in their $\CN=2$ truncation. The near horizon geometry of these  black holes only has $\CN=2$ symmetry so for the purposes of the finding the localizing solutions such a truncation is justified. However, this is a truncation and not a reduction because the masses of the gravitini multiplets that are ignored are of the same order as the scale of the black hole horizon. As a result these multiplets can make a contribution to the the one-loop determinants. Similarly, the hypermultiplets are known not to contribute to the classical entropy but could in principle contribute at the quantum level. The striking agreement with the 
microscopic answers, as well as the macroscopic one-loop determinants~\cite{Sen:2011ba} does suggest that 
the combined effects of the fields that are not kept in the truncation is not significant for our problem at hand. 
To settle this issue definitively one would require an off-shell  realization of the two supercharges used for 
localization but for  all fields  including the additional gravitini and hypermultiplets. 
This is an interesting problem in supergravity. 


 \vspace{2mm}
\noindent $\bullet$ \textit{Localization in supergravity}:
Implementing localization in a theory of gravity is subtle because the choice of the localizing supercharge requires  a background metric. We have treated the metric as any other field in a fiducial background. This introduces an arbitrary background dependence.  It should be possible to implement localization in a background independent way.
 
  \vspace{2mm}
\noindent $\bullet$  \textit{A puzzle}:
The gauge fields in our problem are all abelian. Consequently, the off-shell supersymmetry transformations are quadratic and as a result the quadratic fluctuation determinant of the localizing action around the localizing saddle point would seem to not depend on the collective coordinates $\{ C^{I} \}$ of these instantons. Thus, unlike in analogous problems in nonabelian gauge theories \cite{Pestun:2007rz}, the one-loop determinants appears to 
be independent of the  $\{ C^{I} \}$. This raises a  puzzle. 
The logarithmic correction to the entropy of a black hole of horizon area~$A_{H}$ in $\CN=2$ supergravity 
coupled to~$n_{\rm v}$ vector multiplets has been found to be $\frac{23 -n_{v}}{12} \log A_{H}$ 
from the one-loop contribution the on-shell effective action. In our off-shell analysis, we find, instead 
$-\frac{n_{v} +1}{2} \log A_{H}$. A possible source of  this discrepancy is our  gauge-choice which trades the conformal factor of the metric for a scalar field in the superconformal gravity which can lead to area dependent overall normalization of the degeneracy and thus to a logarithmic contribution to the entropy. 
Investigations on this front, which also addresses the first puzzle about the truncation, are currently 
underway~\cite{MurReys, GupJeon}.

\subsection*{Acknowledgments}

The work of A.~D.~was conducted within the framework of  the ILP LABEX (ANR-10-LABX-63)  supported by French state funds managed by the ANR within the Investissements d'Avenir programme under reference ANR-11-IDEX-0004-02,  and by the project QHNS in the program ANR Blanc SIMI5 of the Agence National de la Recherche.
S.~M.~would like to thank the SFTC for support from Consolidated grant number ST/J002798/1. The research of J.~G.~ has received funding from the European Research Council under the European Community's Seventh Framework Programme (FP7/2007-2013) / ERC grant agreement no. [247252].

\bibliographystyle{JHEP}
\bibliography{kloosterman}

\providecommand{\href}[2]{#2}\begingroup\raggedright\begin{thebibliography}{10}

\bibitem{Sen:2008vm}
A.~Sen, {\it {Quantum Entropy Function from AdS(2)/CFT(1) Correspondence}},
  \href{http://xxx.lanl.gov/abs/0809.3304}{{\tt arXiv:0809.3304}}.

\bibitem{Sen:2008yk}
A.~Sen, {\it {Entropy Function and AdS(2)/CFT(1) Correspondence}},  {\em JHEP}
  {\bf 11} (2008) 075, [\href{http://xxx.lanl.gov/abs/0805.0095}{{\tt
  arXiv:0805.0095}}].

\bibitem{Dabholkar:2010uh}
A.~Dabholkar, J.~Gomes, and S.~Murthy, {\it {Quantum black holes, localization
  and the topological string}},  \href{http://xxx.lanl.gov/abs/1012.0265}{{\tt
  arXiv:1012.0265}}.

\bibitem{Dabholkar:2011ec}
A.~Dabholkar, J.~Gomes, and S.~Murthy, {\it {Localization \&; Exact
  Holography}},  {\em JHEP} {\bf 1304} (2013) 062,
  [\href{http://xxx.lanl.gov/abs/1111.1161}{{\tt arXiv:1111.1161}}].

\bibitem{Banerjee:2009af}
N.~Banerjee, S.~Banerjee, R.~K. Gupta, I.~Mandal, and A.~Sen, {\it
  {Supersymmetry, Localization and Quantum Entropy Function}},  {\em JHEP} {\bf
  02} (2010) 091, [\href{http://xxx.lanl.gov/abs/0905.2686}{{\tt
  arXiv:0905.2686}}].

\bibitem{Banerjee:2008ky}
N.~Banerjee, D.~P. Jatkar, and A.~Sen, {\it {Asymptotic Expansion of the N=4
  Dyon Degeneracy}},  {\em JHEP} {\bf 05} (2009) 121,
  [\href{http://xxx.lanl.gov/abs/0810.3472}{{\tt arXiv:0810.3472}}].

\bibitem{Murthy:2009dq}
S.~Murthy and B.~Pioline, {\it {A Farey tale for N=4 dyons}},  {\em JHEP} {\bf
  09} (2009) 022, [\href{http://xxx.lanl.gov/abs/0904.4253}{{\tt
  arXiv:0904.4253}}].

\bibitem{Eichler:1985ja}
M.~Eichler and D.~Zagier, {\em {The Theory of Jacobi Forms}}.
\newblock Birkh{\"a}user, 1985.

\bibitem{Dabholkar:2012nd}
A.~Dabholkar, S.~Murthy, and D.~Zagier, {\it {Quantum Black Holes, Wall
  Crossing, and Mock Modular Forms}},
  \href{http://xxx.lanl.gov/abs/1208.4074}{{\tt arXiv:1208.4074}}.

\bibitem{Lisa92}
L.~C. Jeffrey, {\it {Chern-Simons-Witten invariants of lens spaces and torus
  bundles, and the semiclassical approximation}},  {\em Commun. Math. Phys.}
  {\bf 147} (1992) 563--604.

\bibitem{Witten:1988hf}
E.~Witten, {\it {Quantum Field Theory and the Jones Polynomial}},  {\em
  Commun.Math.Phys.} {\bf 121} (1989) 351.

\bibitem{Rademacher:1964ra}
H.~Rademacher, {\em {Lectures on Elementary Number Theory}}.
\newblock Robert E. Krieger Publishing Co., 1964.

\bibitem{Dijkgraaf:2000fq}
R.~Dijkgraaf, J.~M. Maldacena, G.~W. Moore, and E.~P. Verlinde, {\it {A Black
  hole Farey tail}},  \href{http://xxx.lanl.gov/abs/hep-th/0005003}{{\tt
  hep-th/0005003}}.

\bibitem{Manschot:2007ha}
J.~Manschot and G.~W. Moore, {\it {A Modern Farey Tail}},  {\em
  Commun.Num.Theor.Phys.} {\bf 4} (2010) 103--159,
  [\href{http://xxx.lanl.gov/abs/0712.0573}{{\tt arXiv:0712.0573}}].

\bibitem{LopesCardoso:1999cv}
G.~{Lopes Cardoso}, B.~de~Wit, and T.~Mohaupt, {\it {Deviations from the area
  law for supersymmetric black holes}},  {\em Fortsch. Phys.} {\bf 48} (2000)
  49--64, [\href{http://xxx.lanl.gov/abs/hep-th/9904005}{{\tt
  hep-th/9904005}}].

\bibitem{Dabholkar:1989jt}
A.~Dabholkar and J.~A. Harvey, {\it {Nonrenormalization of the superstring
  tension}},  {\em Phys. Rev. Lett.} {\bf 63} (1989) 478.

\bibitem{Dabholkar:1990yf}
A.~Dabholkar, G.~W. Gibbons, J.~A. Harvey, and F.~{Ruiz Ruiz}, {\it
  {SUPERSTRINGS AND SOLITONS}},  {\em Nucl. Phys.} {\bf B340} (1990) 33--55.

\bibitem{Maldacena:1999bp}
J.~M. Maldacena, G.~W. Moore, and A.~Strominger, {\it {Counting BPS black holes
  in toroidal Type II string theory}},
  \href{http://xxx.lanl.gov/abs/hep-th/9903163}{{\tt hep-th/9903163}}.

\bibitem{Shih:2005qf}
D.~Shih, A.~Strominger, and X.~Yin, {\it {Counting dyons in N = 8 string
  theory}},  {\em JHEP} {\bf 06} (2006) 037,
  [\href{http://xxx.lanl.gov/abs/hep-th/0506151}{{\tt hep-th/0506151}}].

\bibitem{Sen:2008ta}
A.~Sen, {\it {N=8 Dyon Partition Function and Walls of Marginal Stability}},
  \href{http://xxx.lanl.gov/abs/0803.1014}{{\tt 0803.1014}}.

\bibitem{deWit:1979ug}
B.~de~Wit, J.~van Holten, and A.~Van~Proeyen, {\it Transformation rules of n=2
  supergravity multiplets},  {\em Nucl.Phys.} {\bf B167} (1980) 186.

\bibitem{deWit:1984px}
B.~de~Wit, P.~G. Lauwers, and A.~{Van Proeyen}, {\it {Lagrangians of N=2
  Supergravity - Matter Systems}},  {\em Nucl. Phys.} {\bf B255} (1985) 569.

\bibitem{deWit:1980tn}
B.~de~Wit, J.~W. van Holten, and A.~{Van Proeyen}, {\it {Structure of N=2
  Supergravity}},  {\em Nucl. Phys.} {\bf B184} (1981) 77.

\bibitem{LopesCardoso:1999ur}
G.~{Lopes Cardoso}, B.~de~Wit, and T.~Mohaupt, {\it {Macroscopic entropy
  formulae and non-holomorphic corrections for supersymmetric black holes}},
  {\em Nucl. Phys.} {\bf B567} (2000) 87--110,
  [\href{http://xxx.lanl.gov/abs/hep-th/9906094}{{\tt hep-th/9906094}}].

\bibitem{Mohaupt:2000mj}
T.~Mohaupt, {\it {Black hole entropy, special geometry and strings}},  {\em
  Fortsch. Phys.} {\bf 49} (2001) 3--161,
  [\href{http://xxx.lanl.gov/abs/hep-th/0007195}{{\tt hep-th/0007195}}].

\bibitem{Castro:2008ms}
A.~Castro, D.~Grumiller, F.~Larsen, and R.~McNees, {\it {Holographic
  Description of AdS2 Black Holes}},  {\em JHEP} {\bf 11} (2008) 052,
  [\href{http://xxx.lanl.gov/abs/0809.4264}{{\tt arXiv:0809.4264}}].

\bibitem{Gupta:2012cy}
R.~K. Gupta and S.~Murthy, {\it {All solutions of the localization equations
  for N=2 quantum black hole entropy}},  {\em JHEP} {\bf 1302} (2013) 141,
  [\href{http://xxx.lanl.gov/abs/1208.6221}{{\tt arXiv:1208.6221}}].

\bibitem{Ooguri:2004zv}
H.~Ooguri, A.~Strominger, and C.~Vafa, {\it {Black hole attractors and the
  topological string}},  \href{http://xxx.lanl.gov/abs/hep-th/0405146}{{\tt
  hep-th/0405146}}.

\bibitem{Gomes:2013cca}
J.~Gomes, {\it {Quantum entropy and exact 4d/5d connection}},
  \href{http://xxx.lanl.gov/abs/1305.2849}{{\tt arXiv:1305.2849}}.

\bibitem{deWit:2009de}
B.~de~Wit and S.~Katmadas, {\it {Near-Horizon Analysis of D=5 BPS Black Holes
  and Rings}},  {\em JHEP} {\bf 1002} (2010) 056,
  [\href{http://xxx.lanl.gov/abs/0910.4907}{{\tt arXiv:0910.4907}}].

\bibitem{Sen:2009gy}
A.~Sen, {\it {Arithmetic of N=8 Black Holes}},  {\em JHEP} {\bf 1002} (2010)
  090, [\href{http://xxx.lanl.gov/abs/0908.0039}{{\tt arXiv:0908.0039}}].

\bibitem{Maldacena:1998bw}
J.~M. Maldacena and A.~Strominger, {\it {AdS(3) black holes and a stringy
  exclusion principle}},  {\em JHEP} {\bf 12} (1998) 005,
  [\href{http://xxx.lanl.gov/abs/hep-th/9804085}{{\tt hep-th/9804085}}].

\bibitem{deBoer:2006vg}
J.~de~Boer, M.~C.~N. Cheng, R.~Dijkgraaf, J.~Manschot, and E.~Verlinde, {\it {A
  farey tail for attractor black holes}},  {\em JHEP} {\bf 11} (2006) 024,
  [\href{http://xxx.lanl.gov/abs/hep-th/0608059}{{\tt hep-th/0608059}}].

\bibitem{Maloney:2007ud}
A.~Maloney and E.~Witten, {\it {Quantum Gravity Partition Functions in Three
  Dimensions}},  {\em JHEP} {\bf 1002} (2010) 029,
  [\href{http://xxx.lanl.gov/abs/0712.0155}{{\tt arXiv:0712.0155}}].

\bibitem{Banados:1992wn}
M.~Banados, C.~Teitelboim, and J.~Zanelli, {\it {The Black hole in
  three-dimensional space-time}},  {\em Phys. Rev. Lett.} {\bf 69} (1992)
  1849--1851, [\href{http://xxx.lanl.gov/abs/hep-th/9204099}{{\tt
  hep-th/9204099}}].

\bibitem{Strominger:1998yg}
A.~Strominger, {\it {AdS(2) quantum gravity and string theory}},  {\em JHEP}
  {\bf 01} (1999) 007, [\href{http://xxx.lanl.gov/abs/hep-th/9809027}{{\tt
  hep-th/9809027}}].

\bibitem{Elitzur:1989nr}
S.~Elitzur, G.~W. Moore, A.~Schwimmer, and N.~Seiberg, {\it {Remarks on the
  Canonical Quantization of the Chern-Simons-Witten Theory}},  {\em Nucl.Phys.}
  {\bf B326} (1989) 108.

\bibitem{Sen:2009vz}
A.~Sen, {\it {Arithmetic of Quantum Entropy Function}},  {\em JHEP} {\bf 0908}
  (2009) 068, [\href{http://xxx.lanl.gov/abs/0903.1477}{{\tt
  arXiv:0903.1477}}].

\bibitem{kirk1990}
P.~Kirk and E.~Klassen, {\it {Chern-Simons invariants of 3-manifolds and
  representation spaces of knot groups}},  {\em Mathematische Annalen} {\bf
  287} (1990), no.~1 343--367.

\bibitem{Hansen:2006wu}
J.~Hansen and P.~Kraus, {\it {Generating charge from diffeomorphisms}},  {\em
  JHEP} {\bf 0612} (2006) 009,
  [\href{http://xxx.lanl.gov/abs/hep-th/0606230}{{\tt hep-th/0606230}}].

\bibitem{Dabholkar:2010rm}
A.~Dabholkar, J.~Gomes, S.~Murthy, and A.~Sen, {\it {Supersymmetric Index from
  Black Hole Entropy}},  \href{http://xxx.lanl.gov/abs/1009.3226}{{\tt
  arXiv:1009.3226}}.

\bibitem{Maldacena:1997de}
J.~M. Maldacena, A.~Strominger, and E.~Witten, {\it {Black hole entropy in
  M-theory}},  {\em JHEP} {\bf 12} (1997) 002,
  [\href{http://xxx.lanl.gov/abs/hep-th/9711053}{{\tt hep-th/9711053}}].

\bibitem{Gukov:2003na}
S.~Gukov, {\it {Three-dimensional quantum gravity, Chern-Simons theory, and the
  A polynomial}},  {\em Commun.Math.Phys.} {\bf 255} (2005) 577--627,
  [\href{http://xxx.lanl.gov/abs/hep-th/0306165}{{\tt hep-th/0306165}}].

\bibitem{Castro:2011xb}
A.~Castro, N.~Lashkari, and A.~Maloney, {\it {A de Sitter Farey Tail}},  {\em
  Phys.Rev.} {\bf D83} (2011) 124027,
  [\href{http://xxx.lanl.gov/abs/1103.4620}{{\tt arXiv:1103.4620}}].

\bibitem{Beasley:2009mb}
C.~Beasley, {\it {Localization for Wilson Loops in Chern-Simons Theory}},
  \href{http://xxx.lanl.gov/abs/0911.2687}{{\tt arXiv:0911.2687}}.

\bibitem{Kallen:2011ny}
J.~Kallen, {\it {Cohomological localization of Chern-Simons theory}},  {\em
  JHEP} {\bf 1108} (2011) 008, [\href{http://xxx.lanl.gov/abs/1104.5353}{{\tt
  arXiv:1104.5353}}].

\bibitem{Dabholkar:2005dt}
A.~Dabholkar, F.~Denef, G.~W. Moore, and B.~Pioline, {\it {Precision counting
  of small black holes}},  {\em JHEP} {\bf 10} (2005) 096,
  [\href{http://xxx.lanl.gov/abs/hep-th/0507014}{{\tt hep-th/0507014}}].

\bibitem{Sen:2008sp}
A.~Sen, {\it {U-duality Invariant Dyon Spectrum in type II on T**6}},  {\em
  JHEP} {\bf 0808} (2008) 037, [\href{http://xxx.lanl.gov/abs/0804.0651}{{\tt
  arXiv:0804.0651}}].

\bibitem{Banerjee:2008pu}
S.~Banerjee, A.~Sen, and Y.~K. Srivastava, {\it {Partition Functions of Torsion
  $ > 1$ Dyons in Heterotic String Theory on {$T^6$}}},
  \href{http://xxx.lanl.gov/abs/0802.1556}{{\tt 0802.1556}}.

\bibitem{Banerjee:2008pv}
S.~Banerjee, A.~Sen, and Y.~K. Srivastava, {\it {Generalities of quarter BPS
  dyon partition function and dyons of torsion two}},
  \href{http://xxx.lanl.gov/abs/0802.0544}{{\tt 0802.0544}}.

\bibitem{Dabholkar:2008zy}
A.~Dabholkar, J.~Gomes, and S.~Murthy, {\it {Counting all dyons in N =4 string
  theory}},  \href{http://xxx.lanl.gov/abs/0803.2692}{{\tt arXiv:0803.2692}}.

\bibitem{Rozansky:1994qe}
L.~Rozansky, {\it {A Contribution to the trivial connection to Jones polynomial
  and Witten's invariant of 3-d manifolds. 1.}},  {\em Commun.Math.Phys.} {\bf
  175} (1996) 275--296, [\href{http://xxx.lanl.gov/abs/hep-th/9401061}{{\tt
  hep-th/9401061}}].

\bibitem{Banerjee:2009uk}
N.~Banerjee, I.~Mandal, and A.~Sen, {\it {Black Hole Hair Removal}},  {\em
  JHEP} {\bf 07} (2009) 091, [\href{http://xxx.lanl.gov/abs/0901.0359}{{\tt
  arXiv:0901.0359}}].

\bibitem{Jatkar:2009yd}
D.~P. Jatkar, A.~Sen, and Y.~K. Srivastava, {\it {Black Hole Hair Removal:
  Non-linear Analysis}},  {\em JHEP} {\bf 1002} (2010) 038,
  [\href{http://xxx.lanl.gov/abs/0907.0593}{{\tt arXiv:0907.0593}}].

\bibitem{Denef:2007vg}
F.~Denef and G.~W. Moore, {\it {Split states, entropy enigmas, holes and
  halos}},  \href{http://xxx.lanl.gov/abs/hep-th/0702146}{{\tt
  hep-th/0702146}}.

\bibitem{deWit:2010za}
B.~de~Wit, S.~Katmadas, and M.~van Zalk, {\it {New supersymmetric
  higher-derivative couplings: Full N=2 superspace does not count!}},  {\em
  JHEP} {\bf 1101} (2011) 007, [\href{http://xxx.lanl.gov/abs/1010.2150}{{\tt
  arXiv:1010.2150}}].

\bibitem{Murthy:2013xpa}
S.~Murthy and V.~Reys, {\it {Quantum black hole entropy and the holomorphic
  prepotential of N=2 supergravity}},
  \href{http://xxx.lanl.gov/abs/1306.3796}{{\tt arXiv:1306.3796}}.

\bibitem{Sen:2011ba}
A.~Sen, {\it {Logarithmic Corrections to N=2 Black Hole Entropy: An Infrared
  Window into the Microstates}},  \href{http://xxx.lanl.gov/abs/1108.3842}{{\tt
  arXiv:1108.3842}}.

\bibitem{Pestun:2007rz}
V.~Pestun, {\it {Localization of gauge theory on a four-sphere and
  supersymmetric Wilson loops}},  \href{http://xxx.lanl.gov/abs/0712.2824}{{\tt
  arXiv:0712.2824}}.

\bibitem{MurReys}
S.~Murthy and V.~Reys, ``{Functional determinants, index theorems, and exact
  quantum black hole entropy}.'' In preparation.

\bibitem{GupJeon}
R.~Gupta, Y.~Ito, and I.~Jeon . In preparation.

\end{thebibliography}\endgroup

\end{document}